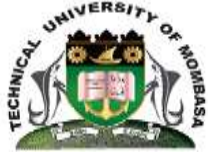

*TECHNICAL UNIVERSITY OF MOMBASA*

*FACULTY OF ENGINEERING AND TECHNOLOGY*

*DEPARTMENT OF ELECTRICAL AND ELECTRONIC ENGINEERING*

**Project report submitted in partial fulfillment requirements for the degree of Bachelor of Science in Electrical and Electronic Engineering**

# Design and Development of an Aerial Surveillance Security System

By

**SIMON KARANJA HINGA**
**kahinga@tum.ac.ke**

**BEEE/013J/2011**

**Supervisor:** MR STEPHEN NABONGO SANDE

MAY 2016

# DECLARATION

I, SIMON KARANJA HINGA, declare that the contents of this project report represent my own unaided work, and that the report has not previously been submitted for academic examination towards any qualification. Furthermore, it represents my own opinions and not necessarily those of the Technical University of Mombasa.

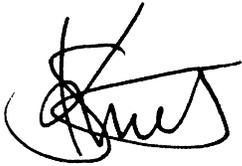

19th May 2016

**Signed**                                              **Date**

# ABSTRACT

Aerial security means performing security-aimed monitoring and surveillance operations with the help of airborne vehicles. This kind of activities suggest that human officers (security organizations, law enforcement, police etc.) would be able to remotely monitor and view video and data acquired from Drones while planning and evaluating their operations. The spectrum of applications where drones are used for security purposes is vast: scouting and reporting emergencies, monitoring accidents and crimes, surveillance of a certain landscape area, operating in highly busy and pedestrians as well as their tracking from up in the sky, and so on.

The project will serve as a bridge to connect actual happening in areas that cannot be navigated easily by security personnel of corporate institution as the Drone will be used to hover and record the actual happening as it transmit to a ground station which records and analyses the events as they streams in, also due its capability of flying over different altitudes the drone can generally be used on areas with rugged terrains or over water bodies for a time dependent on its power capacity.

# ACKNOWLEDGEMENTS

**I wish to thank:**

- The almighty God
- My family, dad, mum, brothers, sisters.
- Irene, Jane and the entire TUM robotics club, and also the B.Sc. EEE class of 2016.
- To my supervisor Mr. Stephen Nabongo Sande.
- To Mr. Kichinda –Technologist, Technical University of Mombasa

# DEDICATION

For

My young nephews JONNY Stive, JONNY Davie, SOLO, MARVIN

My lovely nieces ESSY Stive, ESSY Davie, Karocho MARTHA.

# Contents

# GLOSSARY

**Terms/Acronyms/Abbreviations   Definition/Explanation**

| | |
|---|---|
| **UAV** | **Unmanned Aerial Vehicle** |
| **MAV** | **Micro Air Vehicle** |
| **PDF** | **Payload Directed Flight** |
| **ECMs** | **Electronically Commutated Motors** |
| **BLDC** | **Brushless DC motors** |
| **DC** | **direct current** |
| **AC** | **Alternating Current** |
| **FHSS** | **Frequency Hopping Spread Spectrum** |
| **ESCs** | **Electronic Speed Controller** |
| **IEDs** | **Improvised explosive devices** |

# LIST OF FIGURES

# LIST OF TABLES

# CHAPTER ONE

## 1.1 Introduction

Surveillance is the monitoring of the behavior, activities, or other changing information, usually of people for the purpose of influencing, managing, directing, or protecting them. This can include observation from a distance by means of electronic equipment (such as CCTV cameras), or interception of electronically transmitted information (such as Internet traffic or phone calls); and it can include simple, relatively no- or low-technology methods such as human intelligence agents and interception and aerial surveillance where drones are applied to relay information and gathering the required data.

A quadcopter is an aerial vehicle that uses four rotors for lift, steering, and stabilization. Unlike other aerial vehicles, the quadcopter can achieve vertical flight in a more stable condition. The quadcopter is not affected by the torque issues that a helicopter experiences due to the main rotor. Furthermore, due to the quadcopter's cyclic design, it is easier to construct and maintain. In the project design of a quadcopter is constructed to ensure it can achieve a total flight of 10 minutes in Air with the possibility of future progress to improve in time and robustness.

## 1.2 Overall Objective

To design and construct a working model of a drone that will achieve total autonomous flight.

## 1.3 Specific objectives

1. To perform a literature review on the existing similar systems in order to gain some knowledge that will be applied in implementing this project
2. To design and construct a Quadcopter that can successfully take off, the quadcopter should fly unaided and smoothly.
3. Carry out tests on the designed quadcopter for maximum flight time, maximum height it can fly and the total range of laterally distance it can fly.
4. To implement the actual quadcopter and that will completely carry out instructions and commands given.

### 1.3 problem statement

Visibility is often impaired for those inside the vehicles, making it difficult to see all possible threats ahead, behind, and to the side. Lack of visibility creates a significant danger from insurgents. Travel routes can span hundreds of miles where explosive detectors, bomb-sniffing dogs, or law enforcement is costly, but still does not guarantee complete safety. Improved visibility for individuals in the vehicles can help mitigate these external risks. The prototype can be applied as a bridge that can help record the happening of an area in space giving the real scenario with minimum human guidance

### 1.6 Assumptions and Delimitations

It is assumed that the drone will with stand the effects of wind as its general aerodynamic design is made to cater for any adverse effects. The propellers used by the Drone are designed at an angle that will give 80% efficiency in cases where the wind speed is normal.

### 1.4 Significance and Motivation of Study

This project created a platform to learn about the unmanned aerial vehicles such as the quadcopter. This expands the scope of the Electrical Engineering to include the control and the understanding of the mathematical components. The quadcopter has many applications that an interested to develop security systems, mapping and reconnaissance especially in a disaster and dangerous area. It also opens up the possibilities to broaden the understanding and application of control systems, stabilization, artificial Intelligence and computer Image processing as it applies to the quadcopter.

# CHAPTER 2

**Literature review**

**2.1 Introduction**

This chapter will briefly explain some basic theories on the quadcopters and how the existing security systems utilize quads in the aerial application as Unmanned Vehicle to curb insecurity.

**2.2 Theory and Operation of Quadcopters.**

Manned and large unmanned aircraft of the same type generally have recognizably similar physical components, the main exceptions being the cockpit and environmental control system or life support systems. Some UAVs carry payloads (such as a camera), which aid in taking videos and pictures. Small civilian UAVs have no life-critical systems, and can thus be built out of lighter but less sturdy materials and shapes, and can use less robustly tested electronic control systems. For small UAVs, the quadcopter design has become popular, though this layout is rarely used for manned aircraft. Miniaturization also means that less-powerful propulsion technologies can be used which are not feasible for manned aircraft, such as small electric motors and batteries.

Control systems for UAVs are often different than manned craft. For remote human control, a camera and video link are almost always a necessary replacement for the cockpit windows; instead of physical cockpit controls, commands are received by radio. Both manned and unmanned aircraft can have sophisticated autopilot software, though the features for autonomous drone operation are often different than those for large aircraft such as civilian passenger airliners. The general physical structure of an UAV is as shown below in fig2.1 ("Unmanned Aerial Vehicle" 2016a)

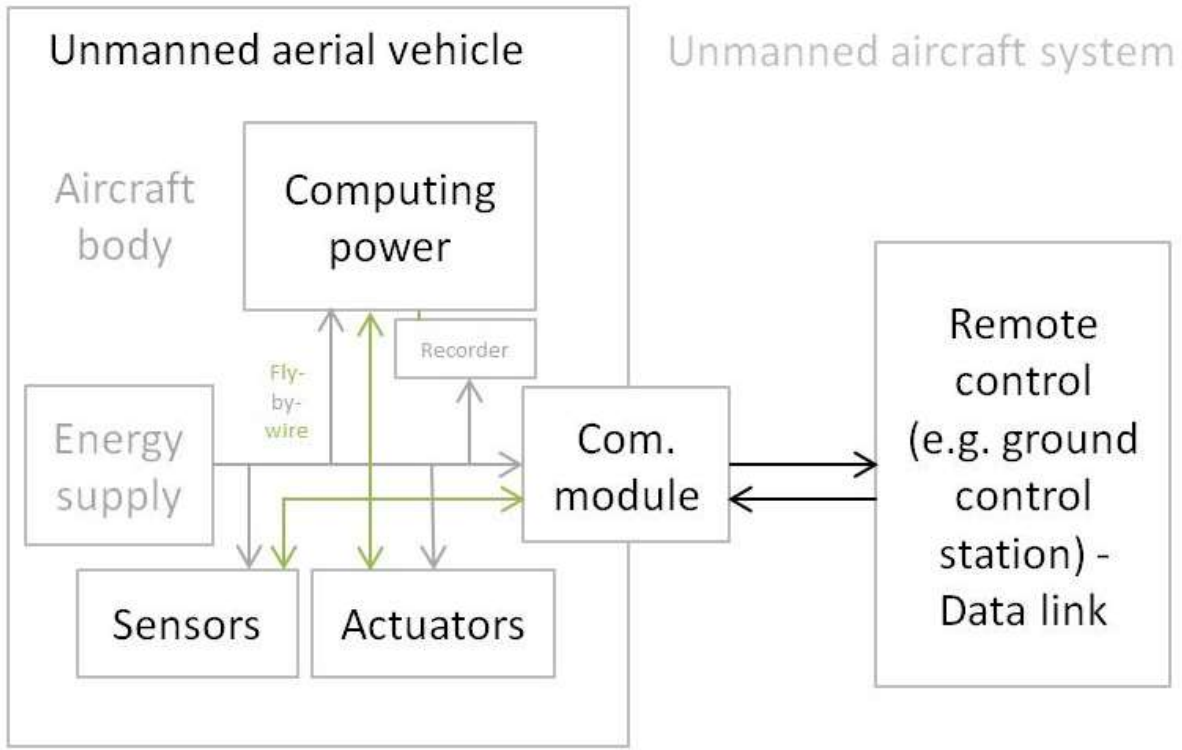

**Figure 2.1 general physical structure of an UAV ("Unmanned Aerial Vehicle" 2016b)**

## 2.3 Power supply and platform

Small UAVs rely mostly at present on lithium-polymer batteries (Li-Po), while larger vehicles often use fuel or even solar power. Battery elimination circuitry (BEC) is used to centralize power distribution and often harbors a microcontroller unit (MCU). Costlier switching BECs diminish heating on the platform.

## 2.4 Computing power

Early large UAVs could carry high computational capabilities due to their extended available payload and they did not urge engineers into miniaturization as they allowed complex-instruction-set chips. Processing power of civil-and-medium-domestic UAVs mostly leans toward reduced-instruction-set computer design. Common processor families there are AVR, PIC, ARM, with a current predominance of ARM's 32-bit memory-address-register processors. Thus, small UAV embedded systems evolved from the blending terms of microcontrollers, to system-on-a-chip (SOC), and as far as single-board computers (SBC) at present. UAV hardware is likely to specialize, with increasing numbers of operation per second and hardware acceleration as a background, between, on one hand, calculus speed in exchange for low processing power (time-critical applications), and high-computational-capacity, able to support full operating systems, trading with higher weight on the other hand. Small UAV control system hardware is often called, especially in hobbyists groups, the Flight Controller (FC), Flight Controller Board (FCB), or Autopilot. ("Unmanned Aerial Vehicle" 2016b)

## 2.5 Sensors

Main sensors:
- Proprioceptive: IMU (gyroscope, accelerometer), compass, altimeter, GPS module, payload measurement...
- Exteroceptive: camera (CMOS, infrared), range sensors (radar, sonar,).
- Exproprioceptive: internal/external thermometer, gimballed camera...

Degrees of freedom (DOF) refer to both the amount and quality of sensors on-board: 6 DOF stands for 3-axis gyroscopes and accelerometers (a typical inertial measurement unit – IMU), 9 DOF refers to an IMU plus a compass, 10 DOF adds a barometer and 11 DOF usually combine a GPS receiver.("Unmanned Aerial Vehicle" 2016a)

## 2.6 Actuators

Actuators found in UAVs depend heavily on the aircraft type: digital electronic speed controllers (which control the RPM of the motors) linked to motors/engines and propellers, servomotors (for planes and helicopters mostly), weapons, payload actuators, LEDs, speakers.

## 2.7 Software

The UAV computer software are layered in tiers with different time requirements. The combination of layers is sometimes called the flight stack, or autopilot.

Onboard classical operating systems alone are not ideal for flying UAVs: high response times may be fatal to the aircraft. Thus they may be completed by externally supported middlewares: RaspberryPis, Beagleboards, etc. shielded with NavIO, PXFMini, etc. or designed from scratch for hard real-time requirements, like Nuttx, preemp-RT Linux, Xenomai, Orocos-Robot Operating System, DDS-ROS 2.0 for instance.

## 2.8 Flight stack overview

| Layer | Requirements | Operations | Examples |
|---|---|---|---|
| Firmware | Time-critical | From machine code to processor execution, memory access. | arduCopter-v1.px4 |
| middleware | Time-critical | Flight control, navigation, radio management. | cleanFlight, Ardupilot |
| Operating system | Computer-intensive | Optic flow, obstacle avoidance, SLAM, decision making | ROS, Nuttx, linux distributions, Microsoft IOT |

**Table2.1: flight stack overview ("Unmanned Aerial Vehicle" 2016a)**

## 2.9 Loop principles

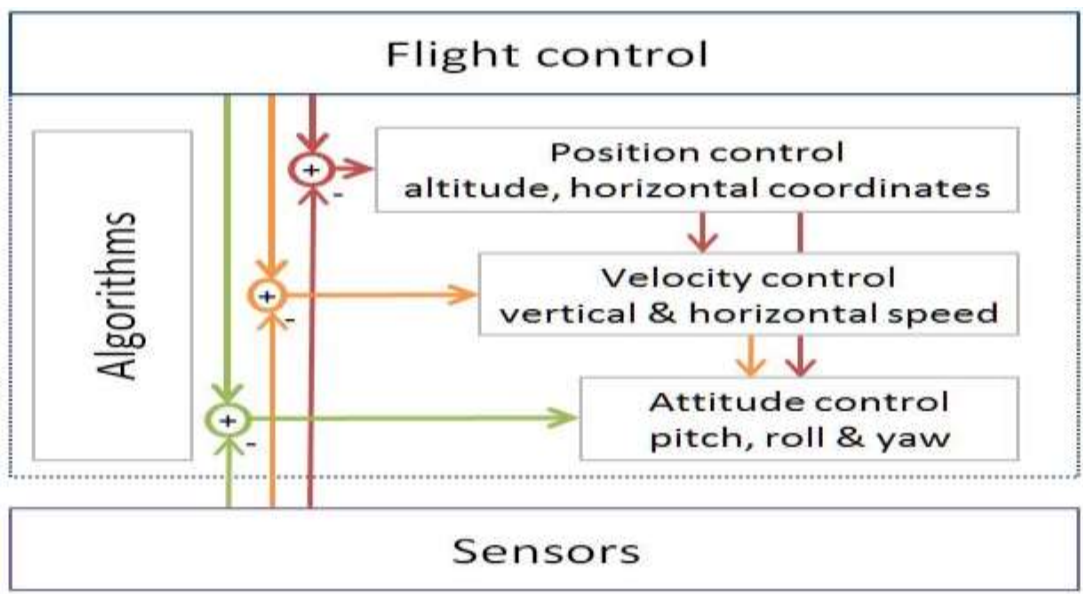

**Fig 2.2 Typical flight-control loops for ("Unmanned Aerial Vehicle" 2016b)**

What may differentiate an UAV from a RC model aircraft is the ability to offer sensing, computing power and automation. The aircraft control falls under the relevancy of control theory, with its associated notions. Indeed, an UAV can make use of different automatic controls, mainly designed with loops:

- Open loops – The simplest design consists of open loops, typically for motors of small UAVs, which are actuated with sheer input, assuming they will perform as expected (though, for many larger aircraft, including UAVs, engine control relies on closed-loops).
- Closed loops – Negative feedback loops use sensors to measure the state of the dynamical system, they are the most commonly used for flight control in UAVs. May use PID control. Sometimes, feed-forward is also employed, transferring the need to close the loop further

## 2.10 Flight controls

Flight control is one of the low-layer systems, and is not much different from manned aviation: plane flight dynamics, control and automation, helicopter flight dynamics and controls, and multi-rotor flight dynamics were in-depth researched long before the rise of UAVs. The automatic flight control is itself layered in multiple levels of priority. UAVs can be programmed to perform aggressive maneuvers or landing/perching on inclined surfaces, even able afterward to climb toward better communication spots, as recently demonstrated by Stanford college.[41] UAVs can also control flight with varying flight modification, such as VTOL designs

## 2.11 Telecommunication system

Most UAVs use an old-fashioned radio frequency front-end, that connects the antenna to the analog-to-digital converter and a flight computer which controls avionics (and which may be capable of autonomous or semi-autonomous operation). Transmission allows remote control of the aircraft and exchange of other data. Early UAVs had only a control uplink, but with progress in embedded electronics, downlinks have been added.("Unmanned Aerial Vehicle" 2016b) In military systems, which drove the duplex communications, and high-end domestic applications, downlink may also convey payload management status and other advanced features. In the domestic-UAV field, the tele-transmission pattern still usually remains as control commands issued from the operator's transmitter (TX) toward the UAV receiver (RX), downstream consisting mainly in analog video content, from the UAV video emitter (VTX) to the operator's video receiver (VRX). Telemetry is another kind of downstream link, transmitting status about the aircraft systems to the remote operator. UAVs use also satellite "uplink" to access satellite navigation.

The radio signal from the operator side can be issued from either:

- A ground control – a human operating a radio transmitter/receiver, a smartphone, a tablet, a computer, or the original meaning of a military ground control station (GCS). Recently control from wearable devices, human movement recognition, human brain waves was also demonstrated.
- A remote network system, like satellite duplex data links for some major military powers. Downstream digital video over mobile network has also entered consumer market recently, while direct UAV control uplink over the cellular mesh is being researched.
- Another manned aircraft or UAV, serving as a nude relay or as a mobile control station - military manned-unmanned teaming (MUM-T) undergoes important research programs

**2.12 Summary**

From the literature review, it is noticeable that previous designs discussed covers the building principles and design of quadcopters; Knowledge learnt here therefore could help design a system that combines both designs. This project therefore fulfills its objectives by the virtual of the possibility of carrying out specified task in the security surveillance as it becomes very relevant in cases of hostage management situation and also as a spy security device from apace and transmit the surveyed information to a base station.

# CHAPTER THREE

**METHODOLOGY 3.1 Introduction**

This chapter discusses the quadcopter design in detail by breaking the design down by components requirement, construction and testing of both hardware and software parts of the project.

## 3.2 Hardware Design

The hardware design focuses on identification of the system's physical components and their interrelationships. Also determines how these components fit into the system architecture. It also discusses the requirements specification of actual hardware and circuit construction.

The tasks involved in design and development of hardware:

- Design considerations.
- Draw the block diagram
- Identify most suitable flight board controller.
- Identify required motors and how to design them.
- Draw the overall circuit diagram

The first thing was to do a design consideration where analysis of the various parts was analysed and the choice for the different components to use are decided.

| PARTS | Number | SPECIFICATION | WEIGHT |
|---|---|---|---|
| Frame | 1 | HJ model | 300g |
| Motors (19g each) | 4 | Model Number – A2212 13T 1000KV | 76g |
| Propellers | 4 | 10x45 | 15g |
| ESC(13g each) | 4 | 30A SimonK model | 52g |
| Battery | 1 | LIPO | 200g |
| Flight Controller | 1 | Hobby king KK2.1 | 55g |
| Connectors | 20 pieces | Deans Ultra Plugs Gold Bullet Connectors | 30g |
| Total | | | 728g |

**Table 3.1 Parts and components.**

In the design of the circuit hardware, the main blocks of the overall system and drawing it as shown in figure 3.1 below.

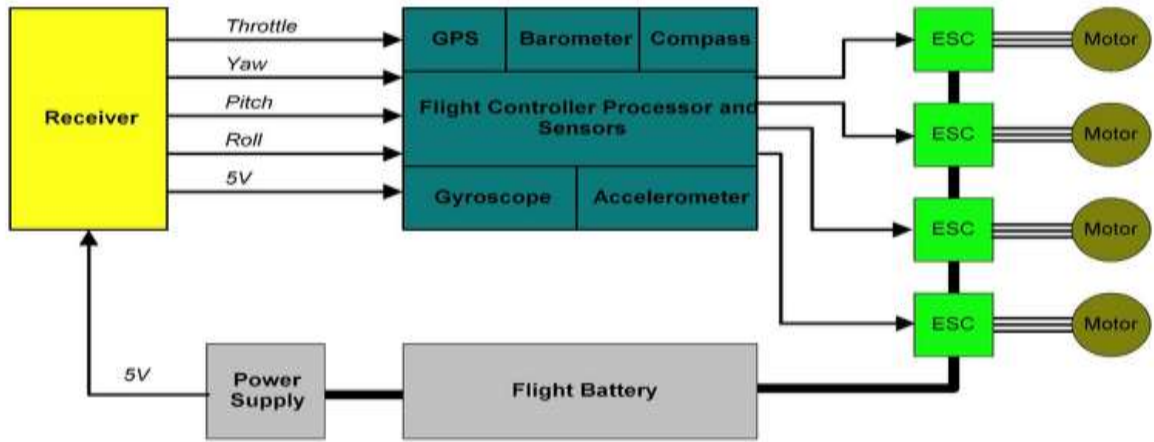

**Figure 3.1 System block diagram** ("QuadcopterBlockDiagram.png (PNG Image, 1026 × 998 Pixels) - Scaled (69%)" 2016)

The third step in the design was to choose a microcontroller suitable for the project implementation. To do this first determine the required hardware interfaces e.g. Flight controllers' module, motors actuators and sensors. The choice of microcontroller depends either on digital or analogue functions. For serial to parallel conversion and vice versa these features are best delivered by the Atmega 644pa in the KK2.1 flight board microcontroller.

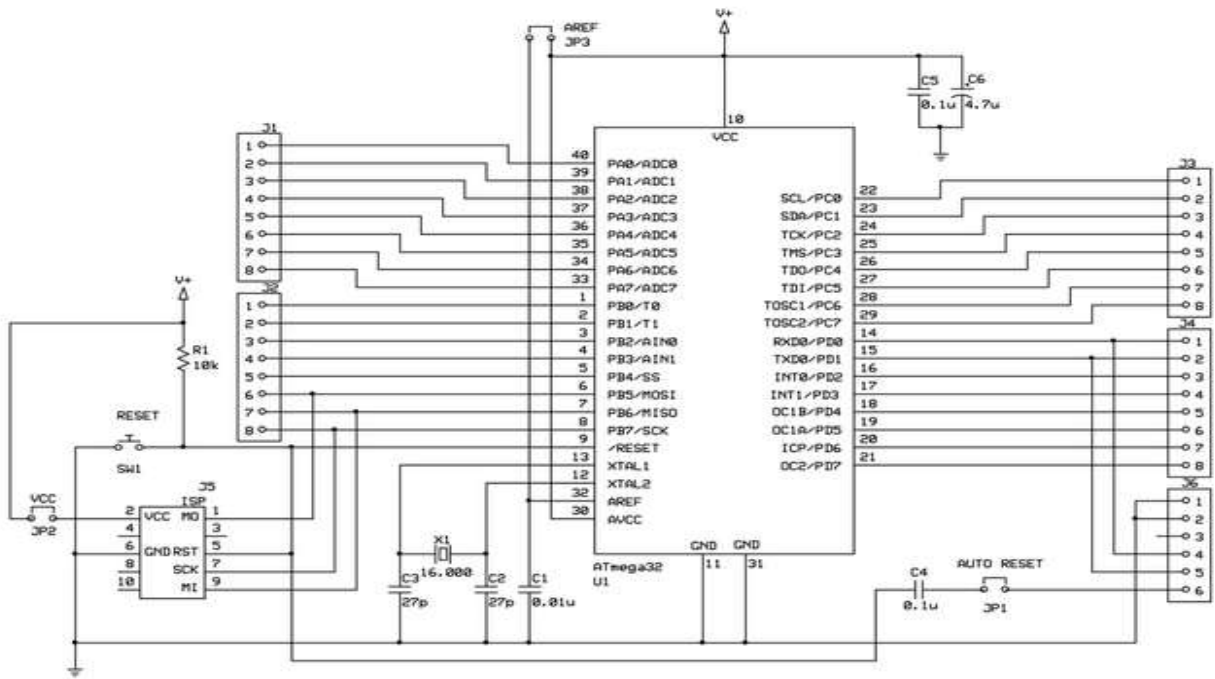

**Fig 3.2 Atmega 644 pa pin connections ("Atmega 40-Pin Development Board avrProgrammers" 2016)**

The forth step in the design process was to design the layout of the motor and electronic speed controller that will be used in the construction as different motor need special circuitry and need battery eliminator circuits that are special for their use.

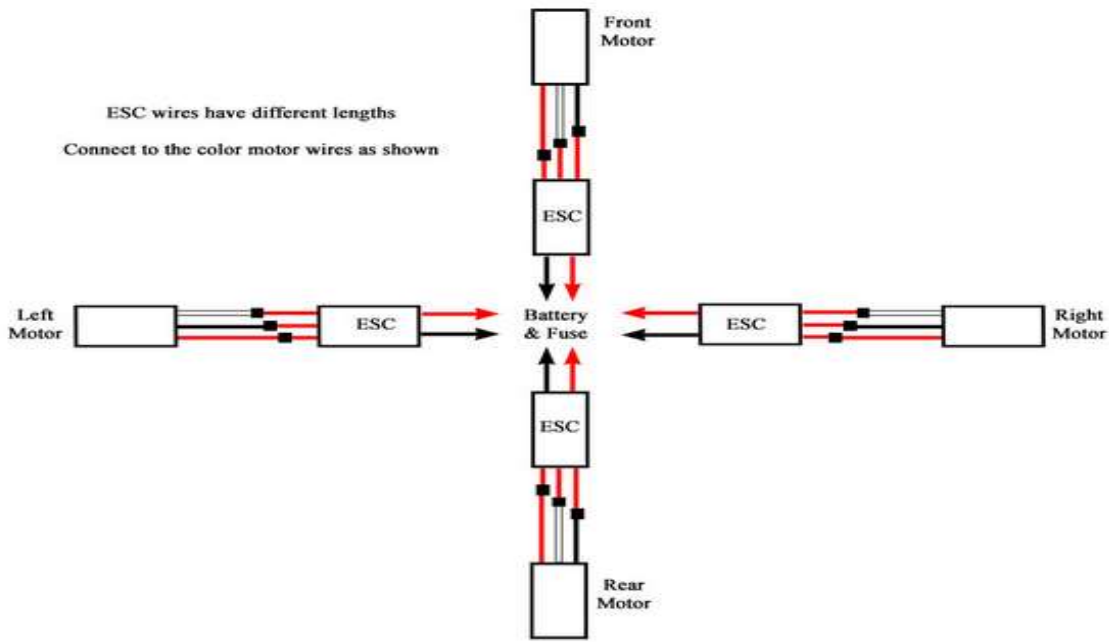

**Fig 3.3 motor and ESC connections (Media365 2016)**

The fifth step in the design process was to determine the type propeller, size and dimensions factoring the aero-dynamics of the design and its ease of manoeuvrability

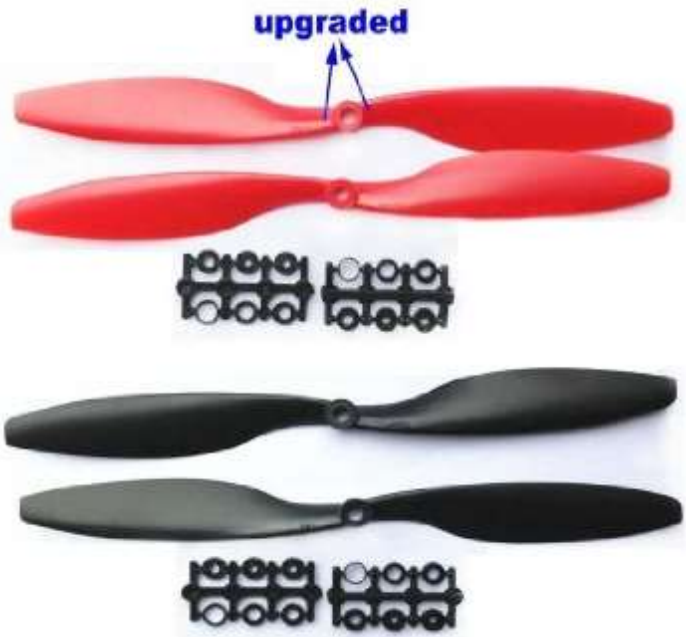

**Fig 3.4 propellers, 2 pairs of red and blue colour each with clockwise and counter clockwise.**

The fifth step was to explain the Quadcopter movement mechanism

## 3.3 Quadcopter movement mechanism

Quadcopter can described as a small vehicle with four propellers attached to rotor located at the cross frame. This aim for fixed pitch rotors are used to control the vehicle motion. The speeds of these four rotors are independent. By independent, pitch, roll and yaw attitude of the vehicle can be control easily. Pitch, roll and yaw attitude off Quadcopter are shown in Figure 3.5, 3.6 and 3.7.

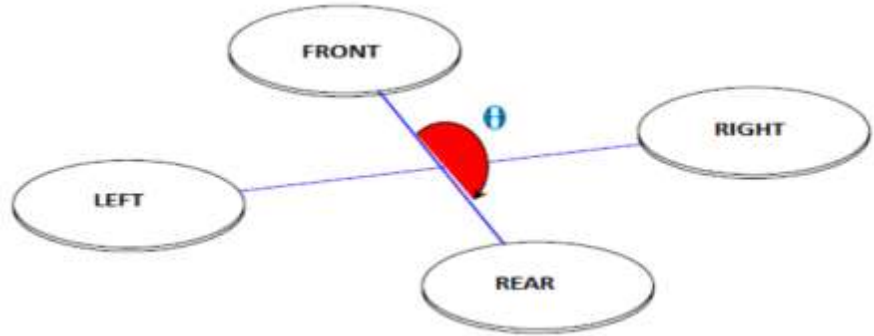

**Fig 3.5 Pitch direction of quadcopter**

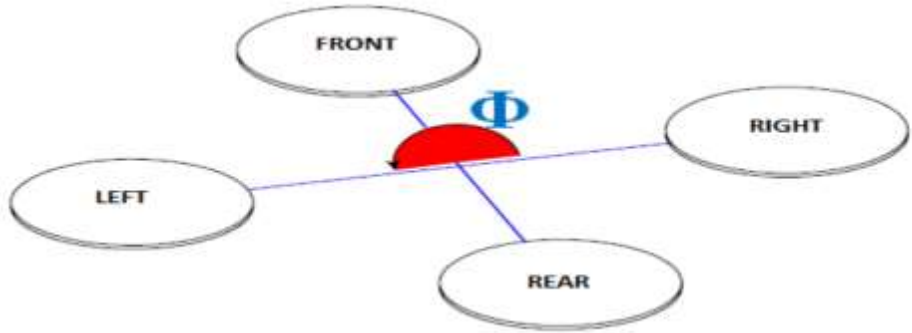

**Fig 3.6 Roll direction of quadcopter**

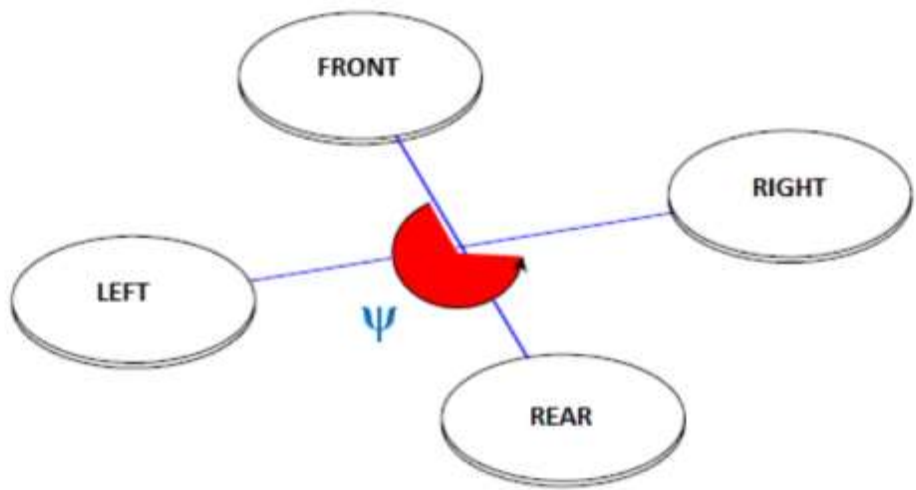

**Fig 3.7 Yaw direction of quadcopter**

Quadcopter have four inputs force and basically the thrust that produced by the propeller that connect to the rotor. The motion of Quadcopter can control through fix the thrust that produced. These thrust can control by the speed of each rotor.

### 3.3.1 Take-off and landing motion mechanism

Take-off is movement of Quadcopter that lift up from ground to hover position and landing position is versa of take-off position. Take-off (landing) motion is control by increasing (decreasing) speed of four rotors simultaneously which means changing the vertical motion. Figure 3.5 and 3.6 illustrated the take-off and landing motion of Quadcopter respectively

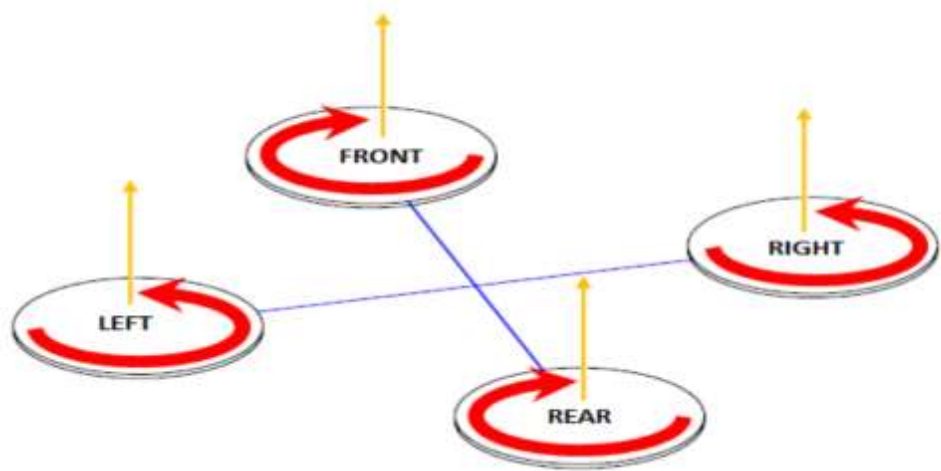

**Figure 3.8: Take-off motion**

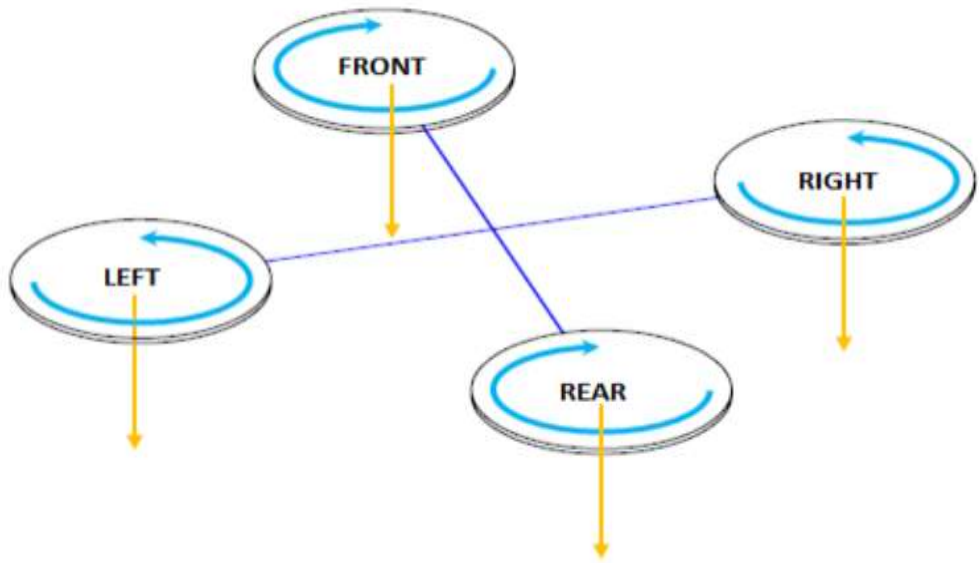

**Figure 3.9: Landing motion**

### 3.3.2 Forward and backward motion

Forward (backward) motion is control by increasing (decreasing) speed of rear (front) rotor. Decreasing (increasing) rear (front) rotor speed simultaneously will affect the pitch angle of the Quadcopter. The forward and backward motions of Quadcopter are represented in Figure 3.7 and 3.8 respectively.

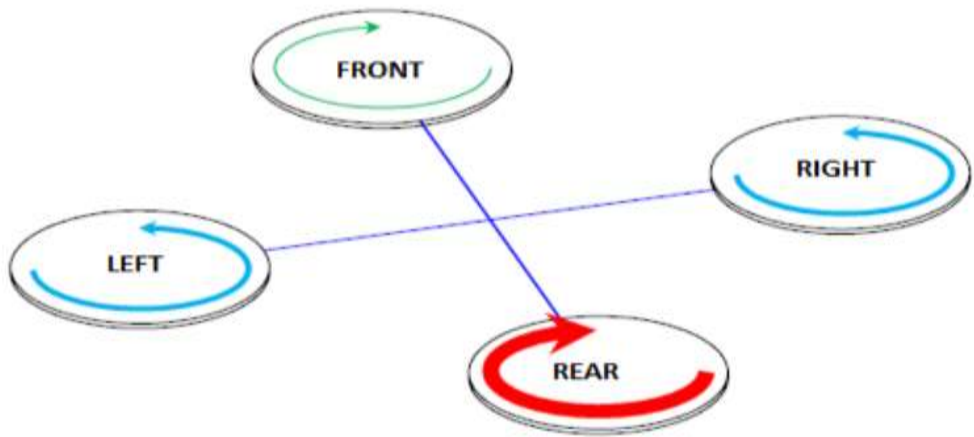

**Figure 3.10: Forward motion**

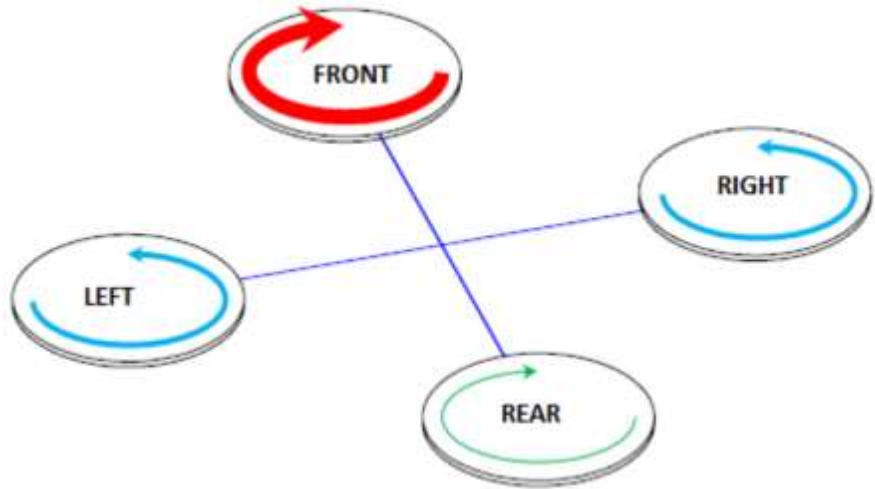

**Figure 3.11: Backward motion**

### 3.3.3 Left and right motion

For left and right motion, it can control by changing the yaw angle of Quadcopter. Yaw angle can control by increasing (decreasing) counter-clockwise rotors speed while decreasing (increasing) clockwise rotor speed. Figure 3.9 and 3.10 show the right and left motion of Quadcopter.

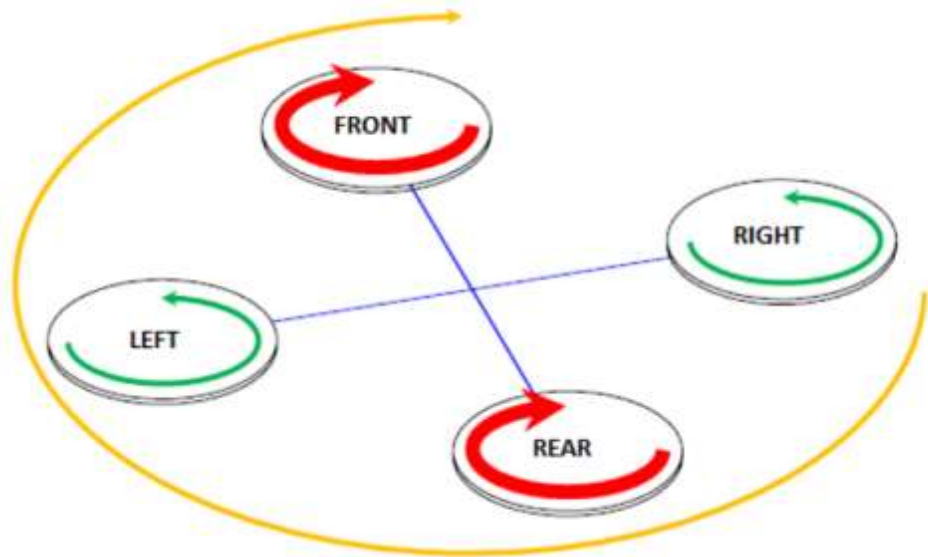

**Figure 3.12: Right motion**

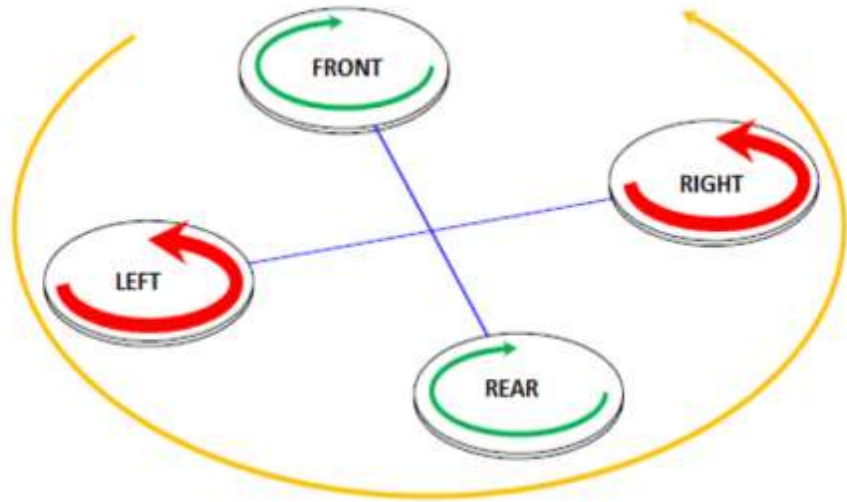

**Figure 3.13: Left motion**

### 3.3.4 Hovering or static position

The hovering or static position of Quadcopter is done by two pairs of rotors are rotating in clockwise and counter-clockwise respectively with same speed. By two rotors rotating in clockwise and counter-clockwise position, the total sum of reaction torque is zero and this allowed Quadcopter in hovering position.

Lastly, the sixth step in the design process is the mathematical modeling of the quadcopter.

### 3.4 Quadcopter mathematical modeling

The schematic movement of Quadcopter is represented in Figure 3.14 and based on schematic, the Quadcopter mathematical modeling is derived as below.

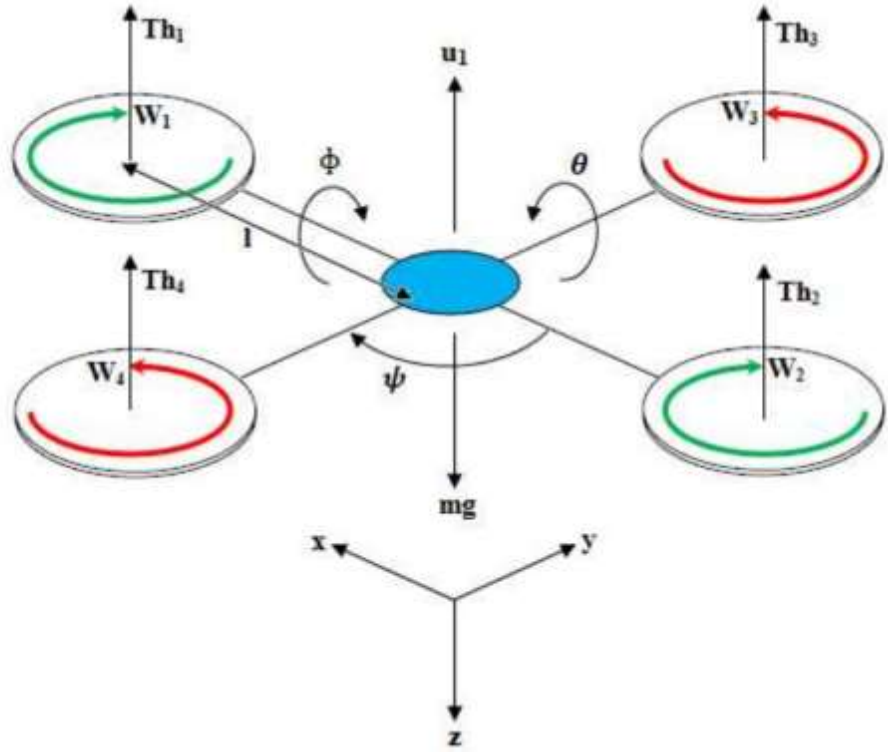

**Figure 3.14: Schematic of Quadcopter**

Where,

$U_1$=sum of the thrust of each motor

$Th_1$=thrust generated by front motor

$Th_2$=thrust generated by real motor

$Th_3$=thrust generated by right motor

$Th_4$=thrust generated by left motor

m=mass of Quadcopter

g=the acceleration by gravity

l=the half length of the quadcopter

x,y,z =three position

θ, ɸ, ψ = three Euler angles representing pitch, roll, and yaw

The dynamics formulation of Quadcopter moving from landing position to a fixed point in the space is given as:

$$R_{xyz} = \begin{bmatrix} C\phi C\theta & C\phi S\theta S\psi - S\phi C\psi & C\phi S\theta C\psi + S\phi S\psi \\ C\phi S\theta & S\phi S\theta S\psi + C\phi C\psi & S\phi S\phi C\psi - C\phi S\psi \\ -S\theta & C\theta S\psi & C\theta C\psi \end{bmatrix} \qquad (3.1)$$

Where,

R = matrix transformation

$S_\theta$ = Sin (θ), Sϕ= Sin (ϕ), S ψ = Sin (ψ)

$C_\theta$ = Cos (θ), Sϕ= Cos (ϕ), S ψ = Cos (ψ)

By applying the force and moment balance laws, the Quadcopter motion equation are given in Equation (3.2) till (3.4) and Pythagoras theorem is computed as Figure 3.15.

$$x = u_1 (Cos\phi Sin\theta Cos\psi + Sin\phi Sin) - K1\dot{x}/m \qquad (3.2)$$
$$y = u_1 (Sin\phi Sin\theta Cos\psi + Cos\phi Sin) - K2\dot{y}/m \qquad (3.3)$$
$$z = u_1 (Cos\phi Cos\psi) - g - K3\dot{z}/m \qquad (3.4)$$

Where,

Ki = drag coefficient (Assume zero since drag is negligible at low speed)

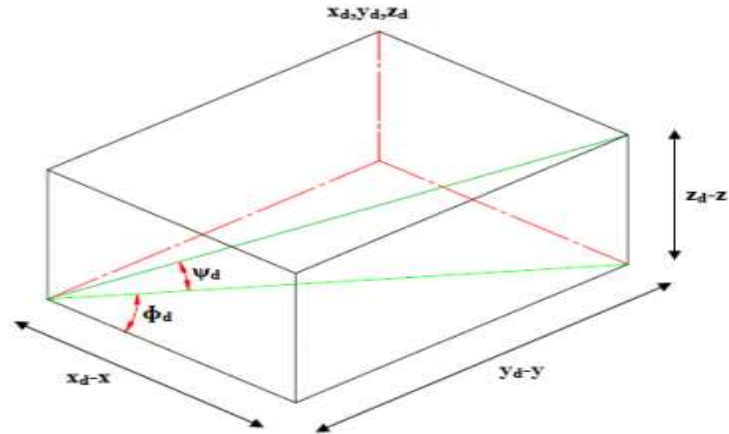

**Figure 3.15: Angle movement of Quadcopter**

The angle ϕd and ψd in Figure 3.12 are determined using Equation (3.5) and (3.6) respectively.

$$\phi_d = tan^{-1}\left(\frac{yd-y}{xd-x}\right) \qquad (3.5)$$

$$\psi_d = tan^{-1}\left(\frac{2d-x}{\sqrt{(xd-x)^2+(yd-y)^2)}}\right) \qquad (3.6)$$

Quadcopter have four controller input forces U1, U2, U3, and U4 that will affectts certain side of Quadcopter. U1 affect the attitude of the Quadcopter, U2 affects the rotation in roll angle, U3 affects the pitch angle and U4 control the yaw angle. To control the Quadcopter movement is done by controlling each input variable. The equations of them are as below:

$$U = \begin{cases} U_1 = (Th_1 + Th_2 + Th_3 + Th_4)/m \\ U_2 = 1(-Th_1 - Th_2 + Th_3 + Th_4)/I_1 \\ U_3 = 1(-Th_1 + Th_2 + Th_3 - Th_4)/I_2 \\ U_4 = 1(Th_1 + Th_2 + Th_3 + Th_4)/I_3 \end{cases} \qquad (3.7)$$

Where,

$Th_i$ =thrust generated by four motor

C=the force to moment scaling factor

$I_i$=the moment of inertia with respect to the axes

Then the second derivatives of each of each angle are:

$$\ddot{\theta} = U_2 - 1K_4\,\dot{\theta}/I_2 \qquad (3.8)$$

$$\ddot{\psi} = U_3 - 1K_5\,\dot{\psi}/I_2 \qquad (3.9)$$

$$\ddot{\phi} = U_1 - 1K_6\,\dot{\phi}/I_3 \qquad (3.10)$$

## 3.5 Hardware design procedure

Here the design is discussed showing all activities involved in framing and implementing the hardware part of the Aerial security surveillance system.
The tasks that will be discussed in this part are:

- Assignment of each motor in the Flight control board and actuators and finally designing the power distribution board the quadcopter.
- Connection of Electronic speed controllers.
- Pairing the radio transmitter to the receiver section to be attached in the Quadcopter.
- Fabrication of the housing to the electronic components and selecting suitable shock-absorbing material.
- Working principles and general assembly of the quadcopter.

**The first step** in the design phase was to assign each motor, actuators and communication devices a pin in the microcontroller. Actuators e.g. buzzer, led and servo motor for the camera gimbal assigned to output pins as shown in the figure 3.16 below

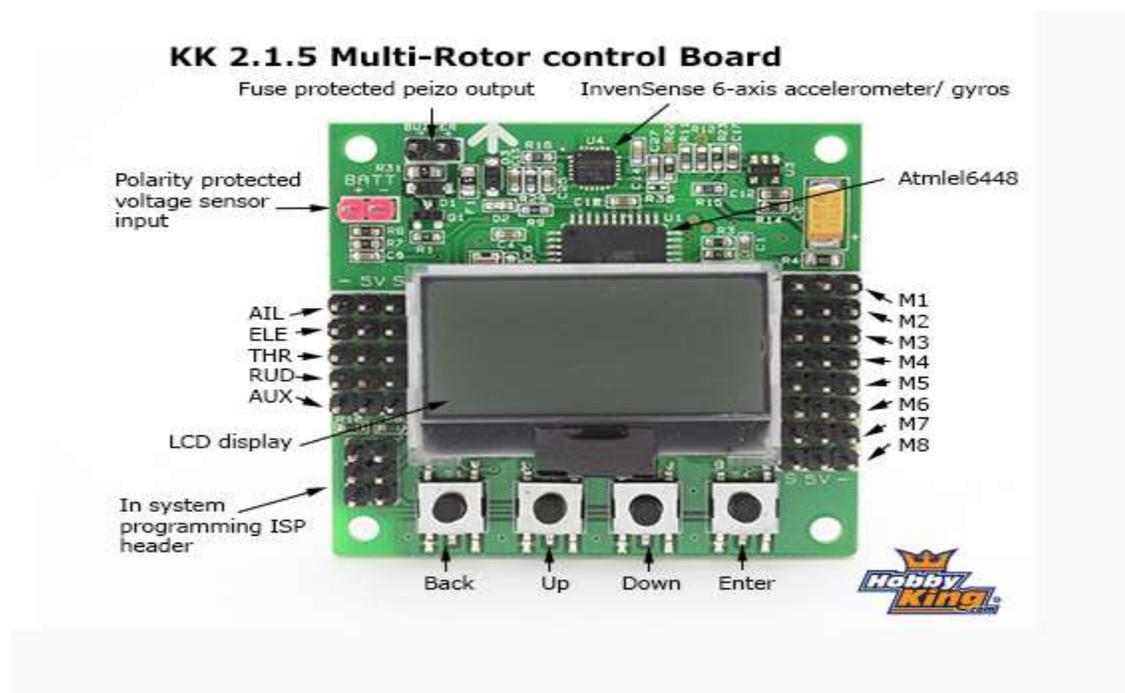

**Fig 3.16 KK2.1.5 flight control board, (hobby king 2015)**

**The Second step** was to connect this ESC to the motor through connecting cables.

### 3.5.1 Connection of the ESC

In order to avoid short circuit and leakage, the joint are connected by thermal shrinkable pipe to ensure Insulation.

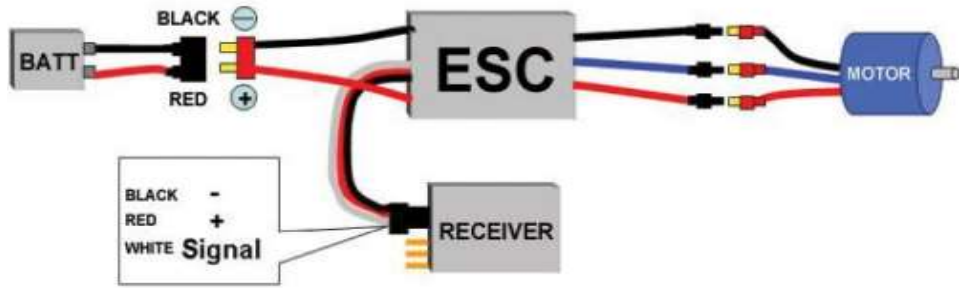

**Figure 3.17 Esc connections to Motors and Battery**

Lastly in construction, fabrication of the housing using 6 inches plywood was constructed as a frame for the quadcopter, the shape design and measurements are as shown in the Appendix. The flight control board was mounted on a soft shock absorbing material that would help protect the gyro the accelerometer, and the Inertia measurement unit.

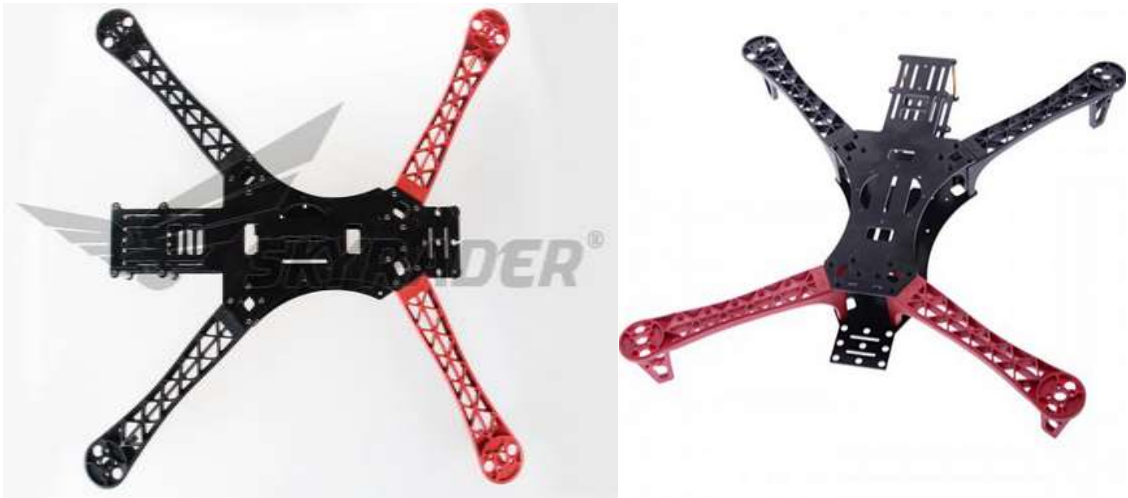

Fig3.18 HJ Quadcopter frame (HJ frame 2016)

### 3.5.2 Circuit Programing

For the quadcopter to work correctly it should be intelligent to take the commands and execute them as guided from the base station which could be miles away from the physical location of the drone. Hence circuit programming is paramount and the heart of the prototype, the flight controller board has four customized buttons and a liquid crystal display screen which aid in the programming. Using Arduino IDE functions which is a higher level programming language is meant for hardware programming.

**Programing algorithm:**

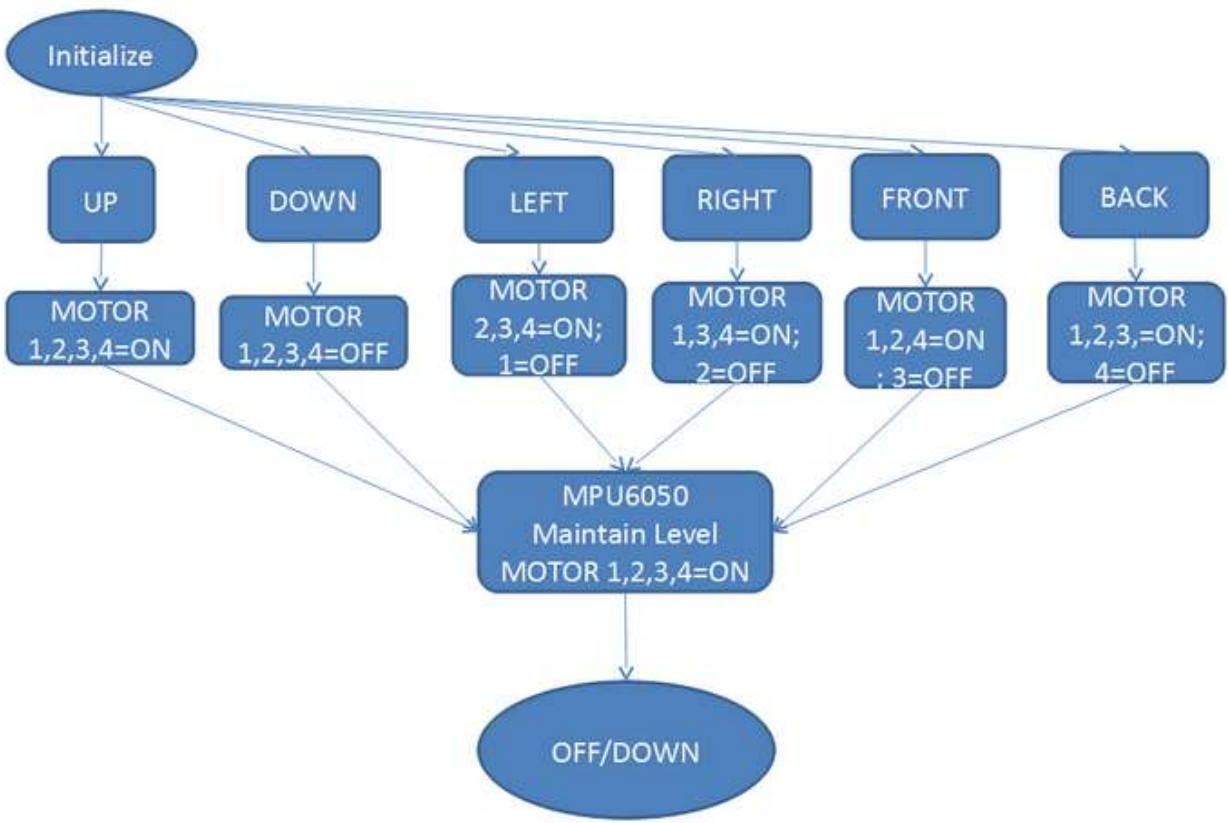

**Fig 3.19 Programming algorithm flow chart.**

## 3.6 Component requirement

Component required are divided into two parts that are hardware and software. Flight controller is applied as auto balance controller of Quadcopter based on input signal from MPU 6050 sensor. The signal produced by KK2.1 Flight controller to control four brushless motor of Quadcopter through Electronic Speed Controller. The Quadcopter body must be rigid and light weight in order to minimize the Quadcopter weigh. For software part, ARDUINO IDE is used to design GUI as interface between control base and Quadcopter.

**Quadcopter Components**

This section discusses the design process behind Quadcopter hardware and software component choices

### 3.6.1 Flight Controller

A flight controller is used to interpret RC controls, provide telemetry data to the base station, as well as provide dynamic control feedback to keep the Quadcopter stable during flight. The flight controller shall support GPS navigation. The flight controller shall use software that can be manipulated by the user. That is, having access to the flight controller code and modifies it per project needs. I applied KK2.1 flight control board.

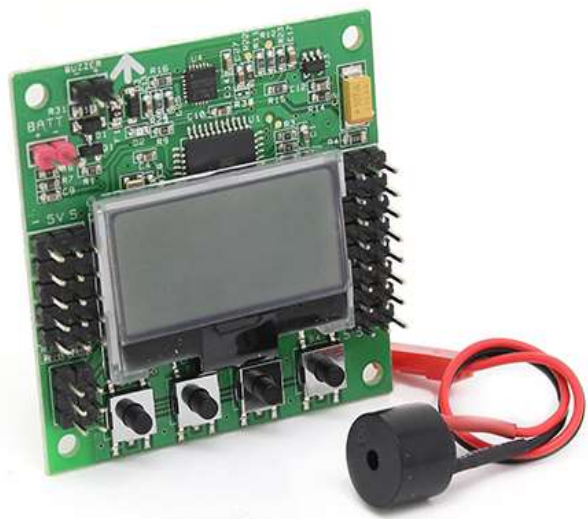

**Figure 3.14 Hobby king KK2.1 Multicolor LCD flight controller board.**

### 3.6.2 Brushless Dc Motors (BLDC Motor)

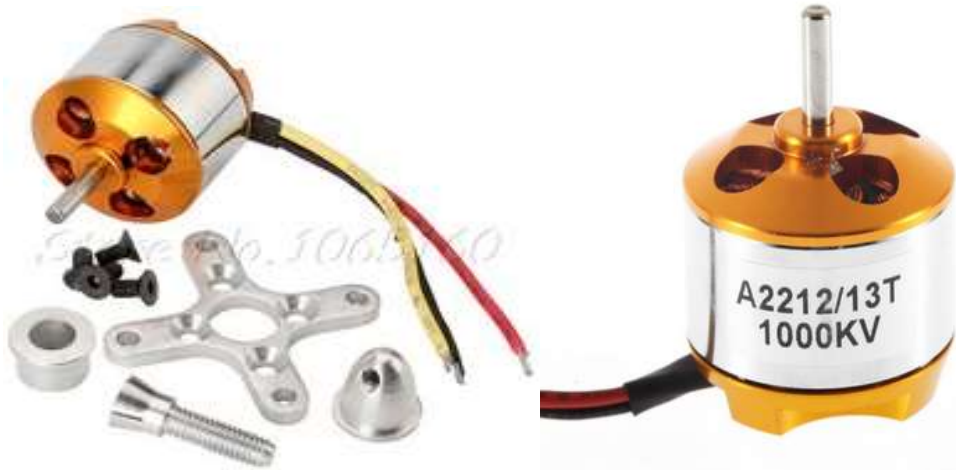

Figure 3.15 brushless DC motor Model A2212/13T (source: hobby king store)

1000Kv Brushless DC Motor is a Brushless DC electric motor (BLDC motors, BL motors) also known as electronically commutated motors (ECMs, EC motors) are synchronous motors that are powered by a DC electric source via an integrated inverter switching power supply, which produces an AC electric signal to drive the motor. In this context, AC, alternating current, does not imply a sinusoidal waveform, but rather a bidirectional current with no restriction on waveform.

 **Working Principal**

A brushless motor is constructed with a permanent magnet rotor and wire wound stator poles. Electrical energy is converted to mechanical energy by the magnetic attractive forces between the permanent magnet rotor and a rotating magnetic field induced in the wound stator poles. There are three electromagnetic circuits connected at a common point. Each electromagnetic circuit is split in the center, thereby permitting the permanent magnet rotor to move in the middle of the induced magnetic field. Most BLDC motors have a three-phase winding topology with star connection. A motor with this topology is driven by energizing 2 phases at a time. The suggested magnetic alignment is used only for illustration purposes because it is easy to visualize. In practice, maximum torque is obtained when the permanent magnet rotor is 90 degrees away from alignment with the stator magnetic field.

## 3.7 ESC (Electric Speed Control)

An electronic speed controller is an electrical circuit that controls the speed of an electric motor and the direction a motor rotates. A motor turns because of the magnetic forces created by the windings and the magnets within the motor.

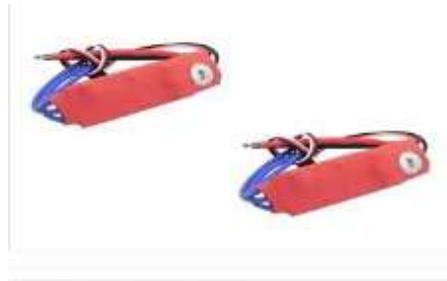

**Figure 3.16 Electric speed controllers**

For a brushless motor, the speed of the motor will depend on the frequency of the winding drive sequence. On a basic brushless motor, there are three windings that are controlled using pulse width modulated (PWM) signals. Two windings will be driven at a time to create the necessary magnetic forces to turn the rotor. The greater the frequency sent to the motors, the faster the rotor will turn due to the magnetic forces. The frequency of the signals is adjusted by changing the pulse width of the signal. Smaller pulse widths will increase the frequency of a PWM signal because more pulses can be sent to the windings in the same time duration, and vice versa for large pulse widths. A brushed ESC works in the same manner but only two control signals are used.

## 3.8 Battery/Power Supply

For our project we need a power supply that is low cost, light weight, reusable, and has enough power for at least ten minutes of flight. Rechargeable batteries were chosen for our project due to reuse value. Currently there are three main types of rechargeable batteries available commercially for radio controlled models, nickel-cadmium (NiCad), nickel-metal hydride (NiMH), and lithium polymer (LiPo) batteries. NiCad batteries have a low internal resistance that allows for high-power output, can operate a large temperature range, but suffers from "memory" loss. This term memory refers to the amount of capacity the battery can store after each discharge. The overall capacity of the NiCad battery will decrease over duration of time.

NiMH batteries are similar to NiCad batteries except they can hold 30% more capacity, but suffer from a larger memory loss.

LiPo batteries can hold 30% more capacity and are much lighter than a NiMH battery. LiPo batteries also suffer from a lower memory loss compared to the NiMH battery. The disadvantages to this battery are that these types of batteries are prone to overheating and overcharging the batteries could lead to fire. Extreme care must be taken when using this type of batteries. Due to these reasons; choose to use a LiPo battery. At this point and time we had just started choosing parts for our quadcopter. To choose the size of battery we had to make some calculations. All calculations and measurements were based on datasheet specifications. To calculate the amount of thrust we needed to overcome gravity, the overall weight of the quadcopter and all its components was found to be about 1550 grams. We may want to add extra components in the future, so we made these calculations with that in mind and assumed the total weight to be1800 grams. Our quadcopter has four motors, therefore 450 grams of thrust was needed for each motor to overcome the forces of gravity.

Using the data sheets I estimated the amount of power needed to run the motors and other components to be about 70 watts. Using the following equation,

$$I = \frac{70w}{11.1v} = 6.3A$$

where I is current in amperes, P is power in watts, and V is voltage in volts. We calculated the amount of current needed for a 3 celled LiPo battery running at 11.1 volts to be about 6.3 amperes. Multiply this by the amount of motors and we needed about 22.2 amperes to fly the quadcopter. Using Peukert's Law, we can determine the capacity amount needed for a estimated flight duration of about 10 minutes.

$$C = \frac{TI^K}{60} = \frac{10MINx22.2A}{60} = 3.7AH$$

C is battery capacity measured in amperes-hour, k is Peukert's constant which we assumed to be 1, and T is time measured in minutes. The calculated capacity per hour was calculated to be 3.7 amperes-hours. I rounded this value up and chose to purchase the Turnigy 3 cell LiPo battery with a capacity of 4000 Mill amperes-hour. All these calculations were based on rough estimates, since I did not have any components available for testing at the time of purchase of the battery. I over specified the weight to make sure I had enough power to fly the quadcopter. The estimated weight was 1500 gram.

This high discharge LiPo is a great way to power any R/C project. This is an excellent choice for anything that requires a small battery with a lot of punch. The discharge rate is high enough to accommodate a lot of electronics and motors. The battery has three cells and outputs 11.1V storing 4000mAh of charge.

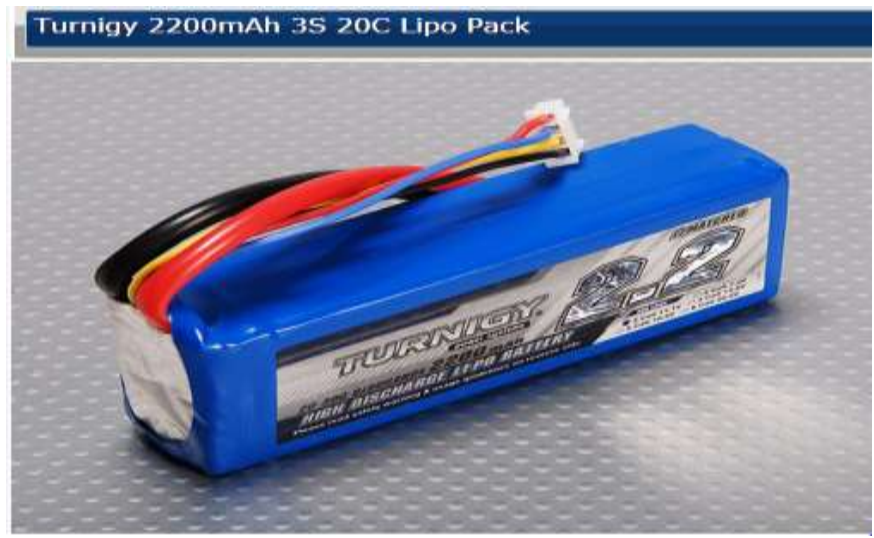

**Figure 3.17 Turnigy 2200mAh 3S lipo Pack**

Specifications

- 11.1 V 3 cell pack
- 4000mAh of charge
- 25C continuous discharge
- Size 103x34x15mm
- Weight 130g

### 3.9 Transmitter and Receiver

I came up with three different options for wireless communication system for our quadcopter. There are three main options to choose from which include a radio transmitter and receiver, a Wi-Fi module, or a Bluetooth module. For my purposes I would need at least six channels to control our quadcopter. The six channels correspond to throttle, roll, yaw, pitch, one channel to switch between acrobatic mode and stable mode, and one or more channels for our auto-commands. Turnigy 9x radio

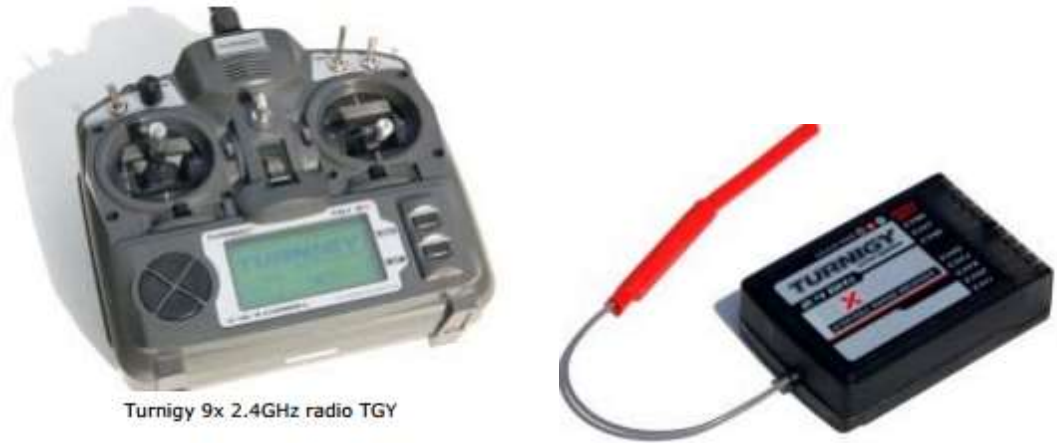

**Fig 3.18 Turnigy 9x 2.4GHz radio TGY.**

**Specifications of Turnigy 9x:**

The 9x TGY is a radio channel dedicated to 2.4 GHz 8, the transmitter specifications

- Encoding PCM 9 channel and 8 channel PPM (2.4 GHz).
- 8 model memories internally with possibility to change their name.
- 167x34mm LCD 8 lines, 22 characters with adjustable contrast.
- Combines the programming for aircraft, glider, helicopter
- Doubles deflections (D / R) and a timer
- Easy access to international grouped on the front.
- Battery Issue 11.1v 1700mAh Lipo up.
- Navigation classic 6 keys.

**Technology:**

The new transmission module (TGY-9X) is of type FHSS (Frequency-hopping spread spectrum) and uses frequency hopping. This technique has high transmission reliability because it is insensitive to interference. His case, plastic is of type 'compact'. Its handling is excellent and his sleeves are adjustable hardness. The antenna is directional. 167x34mm LCD screen (black and white) is fairly well mixed and makes a pretty good readability in daylight. However, the definition of characters and graphics are fairly average Navigation is a conventional type. 6 keys allow access too many menus, selection and settings. The assembly is pretty straightforward except for the fact that the direction of movement of the cursor is not always related to the key position in space. For example pressing a button on the left you can navigate to the right

Transmitter battery: The battery compartment accommodates a holder for 8 AA batteries. It can be retracted to accommodate a liPo battery 11.1v 2500mAh ZIPPY, If you use such a battery, attention to low battery alarm threshold because TGY9x is not set for the li-Po. You will manage your own transmitter battery. If one adds to that a certain lack of precision in the output voltage of the battery issue you should recharge the battery when it reaches the 10s.

Distribution channels on the receiver cannot change the distribution channels on the receiver (this is only possible on top of radio range). Unlike radio Futaba, the path of gas (throttle) does not need to be placed in position 'REV' (reverse).

- Channel 1: Aileron 1 (fins).
- Track 2: Depth (elevator).
- Channel 3: Throttle (throttle).
- Track 4: Management (Rudd).
- Channel 5: Train returning (Gear).
- Channel 6: Shutters (Flaps).
- Channel 7: Auxiliary 1 and fins 2 (for differential).
- Channel 8: Auxiliary 2.
- Channel 9: Battery.

**3.10 Circuit diagram**

## 3.11 Assembly

Assembly begins by attaching a female Dean's ultra-plug to the battery and creating matching pigtail using the male Deans ultra-plug and some short pieces of 12-gauge wire for each the power and ground connections (or alternatively buy a prefabricated pigtail with a male plug). Be careful not to cross-wire this; connect the male and female plugs to verify this. Then solder the pigtail connection to the appropriate terminals of the power distribution mains on the hub by soldering the red wire to positive '+' and the black wire to negative.

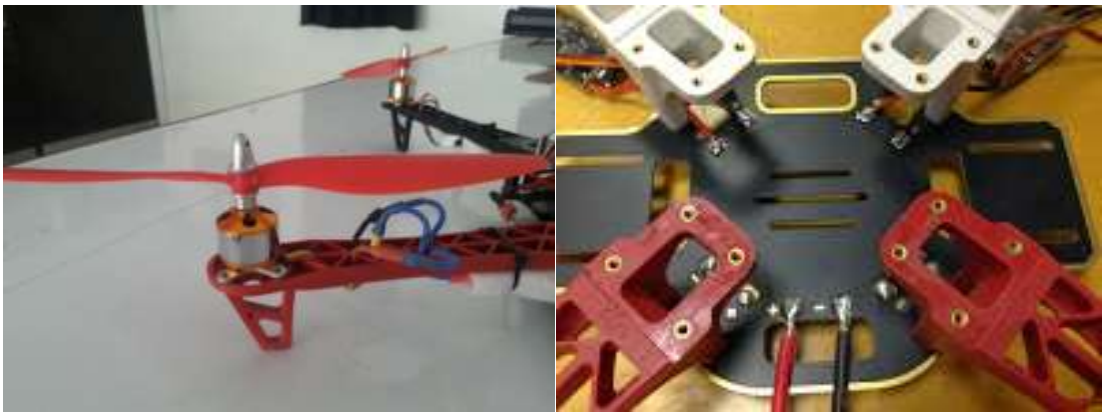

**Fig 3.20 motor connection and frame assembly**

Next, the ESC's are prepared by soldering three female bullet connectors to each of the three (black) wires that will attach the ESC to a motor. Once complete, the arms are attached to the hub (with the white arms at the front and the red arms at the back), the motors mounted to the arms using the screws provided with the flame wheel kit, an ESC is attached to each arm using cable ties and connected to that motor using the bullet connectors. At this time the positive and ground wires coming off the ESC can each be trimmed to length and soldered to the appropriate terminal points on the PDB. For future reference, the side of the quadcopter containing the white arms is designated to be the front and the side with the red arms is the back. This orientation will be maintained for the rest of the assembly and beyond. The use of the red and white arms will further all us to identify the direction that the quadcopter is pointing when it is far away.

At this stage, the assembly is just about complete. All that remains is to add the receiver and the flight controller and wiring it all together (receiver connects to the left side and motors connect to the right side of the KK2 board. The receiver was mounted, using double-sided tape, to one of the side panels. In order to mount the flight controller, I simply used the foam padding container that it was shipped in by hot gluing the controller board to the padding and gluing the padding to the mounting plate on the hub. Using the padding provides the dual benefit of both protecting the flight controller from damage (in the event of a rough landing) as well as reducing the transmissions of vibrations from the frame to the flight controller. Testing of the receiver and the KK2 flight controller can be now be conducted in order to verify that all is working correctly. It should be noted that the KK2 gets its power via the ESC on motor 1. Unless motor 1 is connected (and the battery is attached to the PDB), the flight controller will get no power. Once testing of the KK2 is complete and satisfactory, the top plate of the hub can be attached.

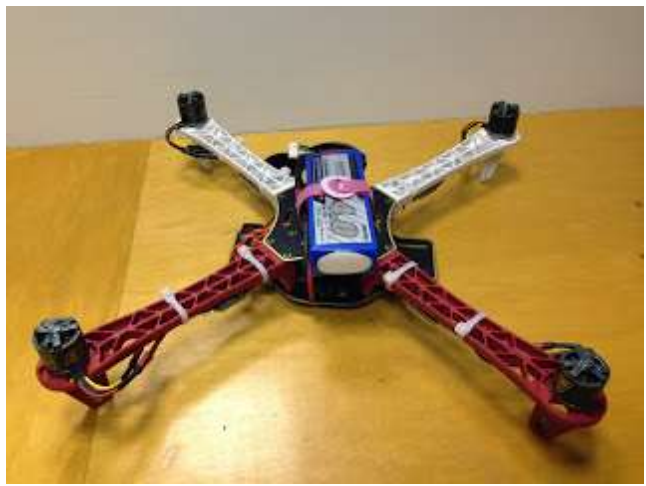

**Fig 3.22 complete layout of the quadcopter**

Before attaching the propellers, it is necessary to ensure that each of the motors is rotating in the correct direction. Each motor is given an identifying number starting with the front left motor which is assigned the number 1 and then continuing from with the front right and then proceeding in a clockwise fashion. It is essential for effective yaw control that diagonally opposite motors spin in the same direction. By convention, motors 1 & 3 should spin clockwise while motors 2 & 4 should spin counterclockwise. In order to spin the motors, we must first arm the KK2. This is achieved by using the left stick in the transmitter and moving it down (zero throttle) and to the right. Holding it there for a couple of seconds will arm the KK2 and increasing the throttle stick will now cause the motors to spin. It should be very quick and easy to verify the direction of rotation of each of the motors. The direction of rotation on any motor can be easily reversed by disconnecting two of the wires between the ESC and the motor and switching them (switching can be done to ANY two wires, it does not matter which to and the motor will reverse - this characteristic of a stepper motor).

The completed receiver and motor wiring of a quadcopter looks as shown in fig 3.23 below

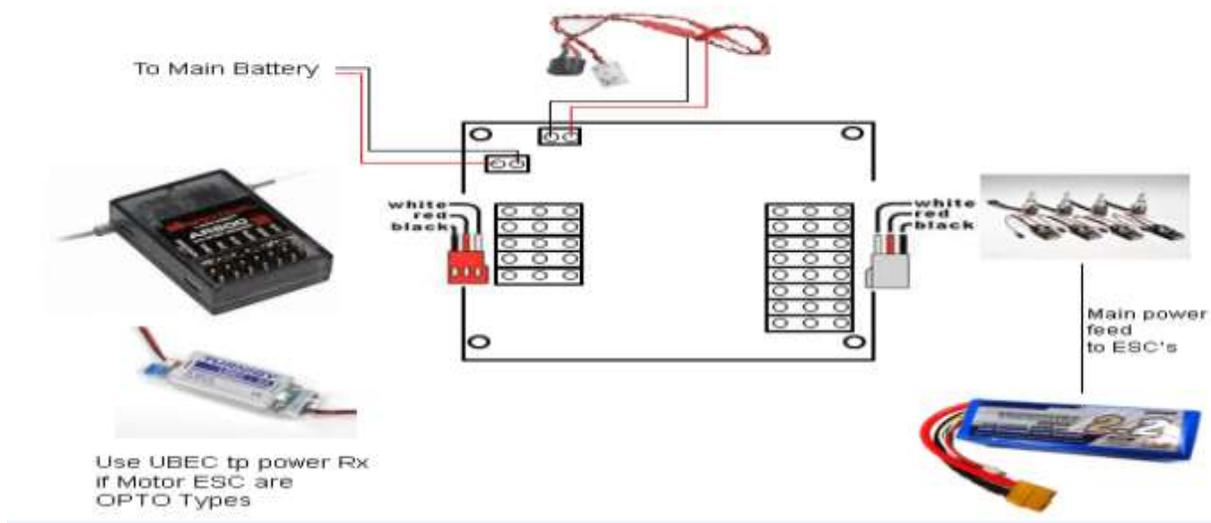

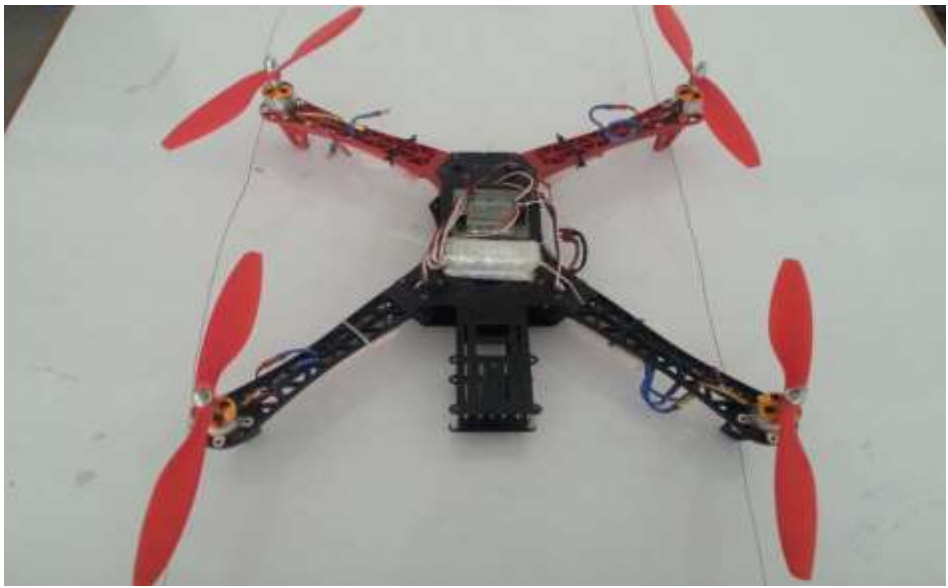

**Fig 3.23 complete wiring of the flight control board, receiver and the motors.**

At this point the quadcopters is ready to mount the propellers and begin tuning the flight controller. First ensure that the gyros are zeroed out on a level surface and will then go about messing with the P and I gains on the controller to ensure that the self-leveling mode is stabilizing the quadcopter's flight. This will be useful for FPV flying and aerial photography while reducing the overall agility of the vehicle.

**3.11 Initial Set-Up of quadcopter before first flight.**

The configuration of the kk2.1.5 is necessary in order to achieve a successful flight, the flight board configuration id normally an important procedure which generally can be viewed as a way of interfacing the different parts of the quadcopter, this was achieved through the following steps, the figure 3.23 below shows the home page of KK2.1.5 while powered.

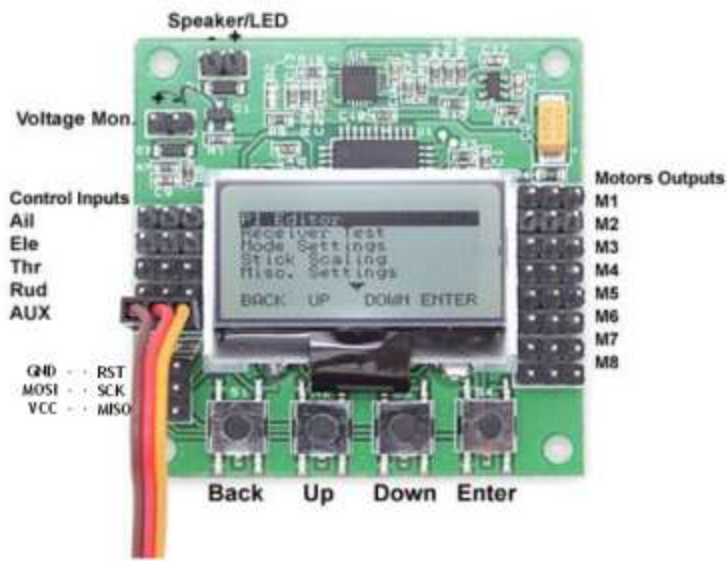

Fig 3.24 KK2.1.5 configuration setup

**STEP-1**

Mount the FC on the frame with the LCD facing front and the buttons facing back. You can use the supplied anti-static foam container as a form of protective case for the Flight Controller on the craft.

**STEP-2**

Connect the receiver outputs to the corresponding left-hand side of the controller board. The pins are defined as

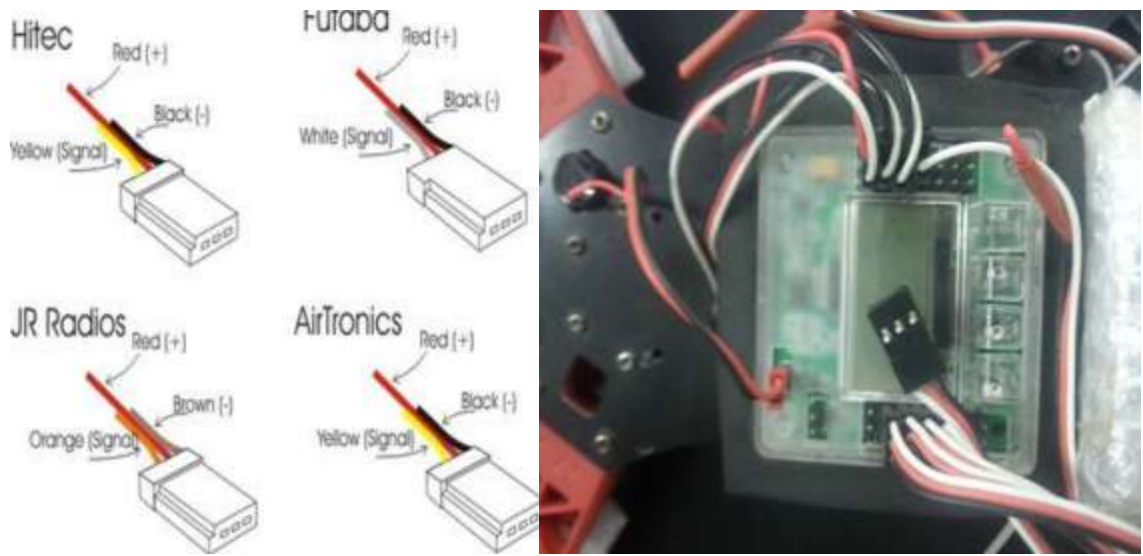

**Fig 3.25 connectors**

Ensure the negative (black or brown) is orientated so that it is on the pin that is nearest to the edge of the Flight Controller Board, so looking at the board the colour sequence will be Black, Red and Orange. The channels are connected as follows from the front of the board towards the push buttons: -

| Receiver channel | Flight control |
| --- | --- |
| Aileron | Aileron |
| Elevator | Elevator |
| Throttle | Throttle |
| Rudder | Rudder |
| AUX1 | AUX |

**TABLE 3.2: receiver channel and flight board connection**

**STEP-3**

Connect the ESC's to the right side of the Flight Controller Board. M1 is towards the front of the board and M4 is nearest to the push buttons. The negative (black or brown) lead towards the edge of the Flight control board. The negative (black or brown) lead is connected to the edge of the Flight Controller.

*Do not mount the propellers at this stage –for safety reasons*

The flight controller Board must always have a source of +V from an ESC, either one of the motors ESC or a separate unit feeding the receiver. If each ESC has a BEC (normal unless OPTO types) then it may be necessary to remove the power feed from the other ESC, usually by cutting the power line (RED) cable on the other ESC.

**STEP-4**

Set up the model on the transmitter and use a normal airplane profile and bind the receiver to the transmitter.

**STEP-5**

Turn on the power and press the 'Menu' button, then using the 'Up' and 'Down' buttons highlight 'Receiver Test sub-menu and press Enter .Now move each channel on your transmitter and check that the displayed direction corresponds with the stick movements on the Flight Controller, if any are reversed, then go to your Transmitter and reverse that channel.  Check that the AUX channel is showing "ON" when you activate the AUX Switch on your transmitter, if not, reverses the AUX channel on your transmitter. Use the trim or sub-trim controls on your transmitter to adjust the channel values shown on the LCD to zero.

**STEP-6**

Scroll down to and enter the "Load Motor Layout" sub-menu and choose the configuration you want. If the configuration you want is not listed, use the "Mixer Editor" sub-menu to make one. See later for more on that.

**STEP-7**

Enter the "Show Motor Layout" sub-menu and confirm the following. load the configuration as x-model Quadcopter.

**STEP-8**

Enter the "receiver test" and check for normal values on each channel, move your transmitter sticks around to ensure they are all working, including AUX1

**STEP-9**

Enter the "PI Editor" sub-menu and check PI gain values using this portion to adjust gain settings. The PREV and NEXT buttons to select the parameter to change, then press change to adjust both Roll and Pitch at the same time, see the "mode setting" submenu. At this stage the propeller can be fitted to test the flight control board, hold the craft and Arm with the right Rudder and zero throttle for a few seconds, it will beep and the RED LED will turn on. Usually you should not arm it until you have put the Quadcopter on the ground and stepped away, after landing, place it in safe Mode by holding the Rudder to the Left with zero throttle, it beeps and the RED LED will turn off, always do this before you approach the quadcopter.  If the craft wants to tip over right away, check the connections and your custom made mixer table if you have one .If it shakes and climbs after it's airborne, adjust the Roll and Pitch P-gain down or if it easily tips over after it's airborne, adjust up. If it drifts away, use the trims to keep the drift down. It will normally drift with the wind. If you need excessive trim, check if the arms and motors have the correct angles and that the motors are good. Increase the Roll and Pitch I gain (note the difference from P gain) until it flies straight forward without pitching up or down. Turn on the Self- leveling by holding right aileron while arming or disarming it. Turn it off by holding left aileron. Alternatively you can assign this to the AUX channel. See below so Sub-menu descriptions.

**STEP-10**

Enter the "Mode Settings" and check and adjust: "Self-Level": Determines how the self-leveling function will be controlled, either by STICK or an AUX Channel. "STICK MODE": Self - Leveling is turned on by holding the aileron to the right when arming or disarming. Turn it off with left aileron. "AUX": Self-leveling is turned on/off by the AUX Channel." Auto Disarm ": If set to YES then Flight Control board will automatically disarm itself after 10-mins of in activity "CPPM Enabled": Determine if the Flight Control Board is to use CPPM data input.

**STEP-11**

Enter the "Stick Scaling" option, where you can adjust the response from the stick to your liking. Higher number gives higher response and lower numbers the converse. This is similar to the endpoint or volume adjustment on your transmitter, where you can adjust your transmitter to adjust the stick response and use the stick scaling if you want more or less Response from stick inputs. "Misc. Settings": "Minimum Throttle": Adjust the setting so that the motors just keep running when the Transmitter throttle stick is at a minimum. "Height Dampening": This option uses the Z accelerometer to dampen vertical movements caused by wind or when tilting the craft. A recommended setting is 30. "Height D. Limit ": Adjust to limit control for Height Dampening to prevent over control, this limits how much power is available for dampening A recommended setting is 10 (10%)." Alarm 1/10 volts": Adjusts the battery alarm voltage set-point. When set to 0 (zero) the alarm is disabled. Adjust this value to suit the battery in use and monitored by the Flight Control Board sensor input. For a standard 3-cell LiPo battery of 11.1volts use a value of 3.60 volts per cell to denote an empty battery and then set this value (in 1/10's) to (3.6 x 3 * 10) = 108 and when the supply voltage drops to 10.8volts the alarm will sound. Note, if you set this value above zero and no battery is attached / monitored then the alarm will sound. As the voltage being monitored nears the set point the time between beeps will shorten, so a long time between pulses when the alarm voltage is getting close to very short time intervals when the voltage is at the alarm set point. "Servo Filter": This setting is a Low-Pass Filter that enables channel jitter to be ignored; a good setting to start off with is 50(mS). If you experience channel jitter then increase this value, if none then set to 0 (zero).

**"Sensor Test":** Displays the output from the sensors. See if all shows "OK". Move the FC around and see that the numbers change.

**"ACC Calibration":** Follow the instructions on the LCD to calibrate the Acceleration Sensors, which is only necessary to do once at initial setup

### 3.12 Test analysis and conclusion.

During design of the prototype, some test points were considered as a measure to give the actual response of the quadcopter. The tests were done based on several steps as outlined below:
  I. Receiver test
  II. Time and duration of flight
  III. Destructive test and Design Material.
  IV. Vibration test

**STEP I. Receiver Test**

It displays the receiver signal inputs.
   a) Use the transmitter trims to set the Roll, Pitch and Yaw values to zero.
   b) Ensure the Throttle is 0 and says "Idle" at low throttle and at full throttle, it is greater than 90 and says "Full". Adjust transmitter throttle trim for low throttle and end point for high throttle.
   c) Roll, Pitch and Yaw should all read between -100 to -90 and 90 to 100 at maximum stick travel. Adjust transmitter end points to achieve this. Do not exceed +/-110.
   d) Ensure Roll, Pitch and Yaw stick commands are correctly shown as Left, Right, Forward, back. If not, reverse the throws in your transmitter.
   e) Arm Test will normally show "Safe Zone". At minimum throttle (throttle 0) and full right yaw, it should display "Arm". At minimum throttle (throttle 0) and full left yaw, it should display "Disarm". Providing there are no ERRORS on the SAFE screen, your multicopter should arm and disarm.
   f) If "No signal" is displayed, check connection to the receiver. Also ensure your receiver is working with your transmitter by connecting a servo to a spare receiver output.

g) Check the Auxiliary channel input and reverse the channel in your transmitter if necessary.

h) Do not use dual rates on your transmitter. Use Stick Scaling instead. This is very important on the Yaw channel as a low rate on the Yaw will prevent Arming and Disarming.

i) If the receiver values appear random, check the following: -
    - Receiver connection(s).
    - Mode Settings, Receiver is correct.
    - If Mode Settings Channel Map is "Yes", check Receiver Channel Map

**STEP II Time and Duration of Flight**

Outdoor test were done were time and distance was recorded showing the maximum distance the quadcopter was able to stay in air.

Table 3.3 Flight test results

| DAY | FLIGHT NUMBER | MAXIMUM TIME (minutes) | HORIZONTAL DISTANCE(meters) | VERTICAL HEIGHT (Meters) |
| --- | --- | --- | --- | --- |
| Day 1 | Test1 and 2 | 3 | 10 | 1 |
| Day 2 | Test 3,4 and 5 | 4.5 | 30 | 2.5 |
| Day 3 | Test 6 | 5 | 45 | 6 |
| Day 4 | Test 7 and 8 | 8 | 56 | 12 |
| Day 5 | Test 9 and 10 | 8.5 | 105 | 18 |

The graph below represents the different flight test conducted during an interval of eight different days, the time of flight was recorded against the total horizontal distance that the quadcopter managed to maneuver laterally and also the approximate vertical height and recorded as shown on the figure above, after the test, the values obtained were interpreted on a bar-graph and a line chart shown the behavior of the quadcopter.

The graph 3.26 below represent the quadcopter horizontal distance attained under the test taking the time and lateral distance.

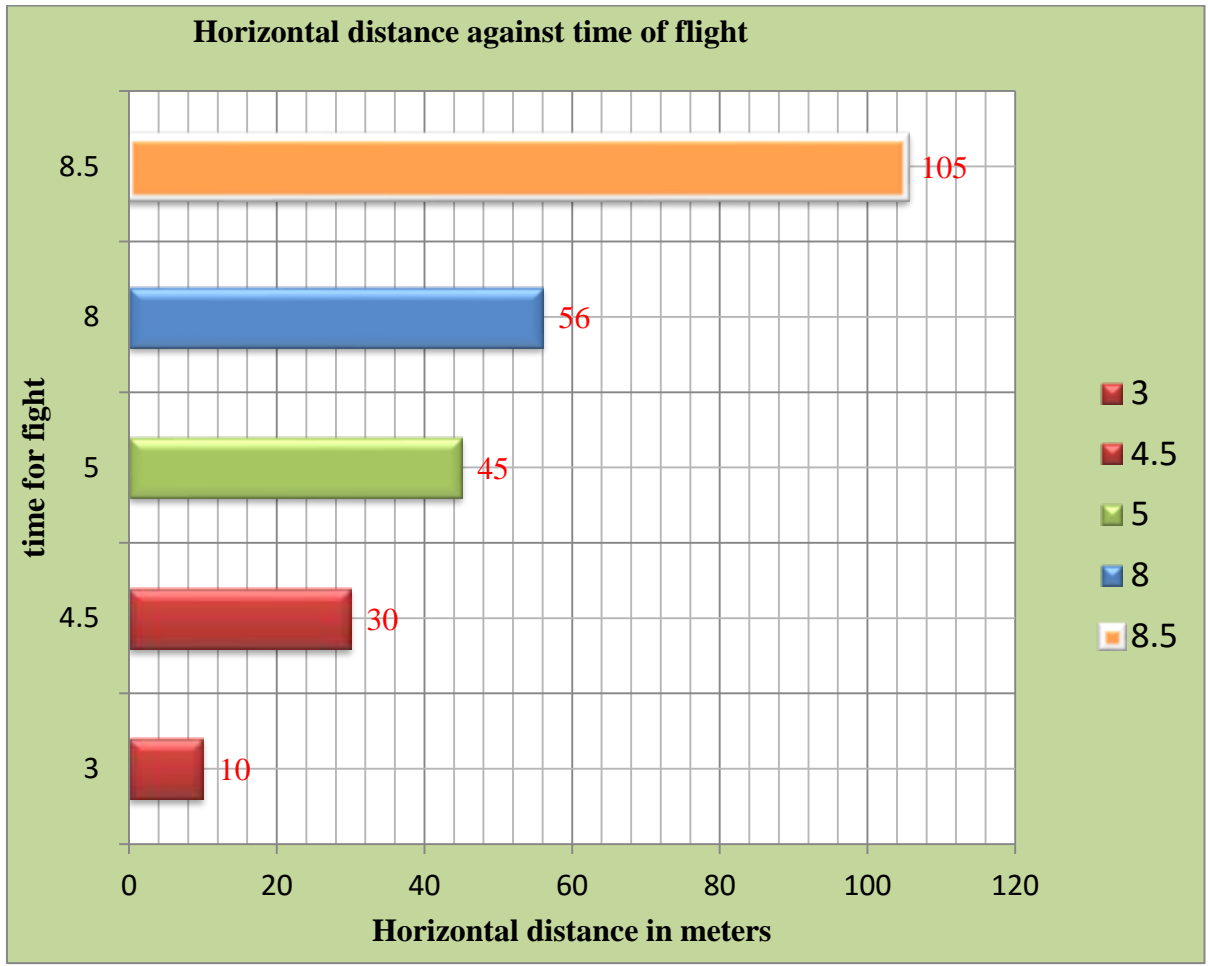

**Figure3.26 horizontal distance against time of flight**

The figure 3.27 below represent the quadcopter horizontal distance attained under the test taking the time and lateral distance

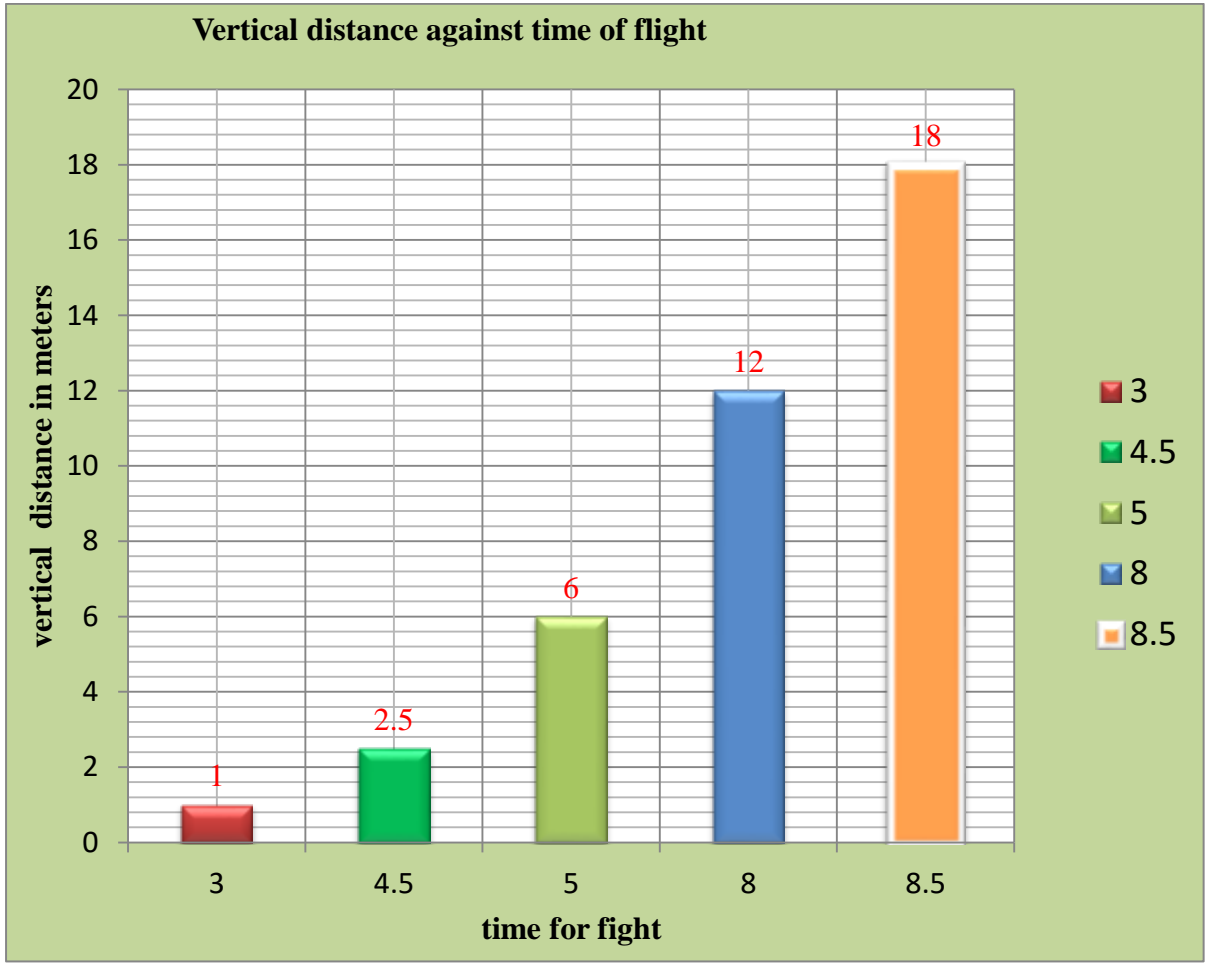

**Figure3.26 horizontal distance against time of flight**

**STEP III. Destructive test and robustness.**
These was the hardest and most expensive test as the test included intentional collision of the quadcopter on objects and making a crash from a height of 10Meters to test the compactness and rigidity of the material used, initial design included using a ply wood made frame with the measurement attached in the appendix, in the material did not withstand the test and they did broke in to pieces, ply wood had the advantage of light weight therefore reducing on the overall weight, the second material used was aluminum made of the same measurement as the ply wood, the aluminum had a shortcoming due to increased vibration with had an adverse effect on the Gyroscope used in the flight board as it made the quadcopter hover a lot. Finally further test was done using the HJ Frame made of carbon fiber and aluminum nuts. This frame was the most stable in its design and Aerodynamics design as it was well balance and gave the best motor angles thus giving a very smooth flight. In destructive the quadcopter was fitted with the HJ Frame and flight of 15metres vertically achieved, the quadcopter was crashed, after the test the propeller broke down, the frame arm did break but the rest remained compact.

**STEP V-Vibration test**

During the design stages of testing, the quadcopter flew for a short period in order to verify it was stable and controllable, during these tests the landing gear was able to handle small falls and minor hits without damaging the Quadcopter. After major fall however, the landing gear induced vibrations that affects the gyro performance causing lateral motion thus disturbing the quadcopter motor, also the setting on PID caused a lot of vibration while the values of the pitch gain and roll values. This was solved by attaching shock absorbing materials on the landing gear to absorb the vibrations resulting.

### 3.13 Challenges, Conclusion and Recommendations

**Challenges**

1. The components were too expensive, the estimated cost was around twenty thousand but it was more than this value due to some un-foreseen expenses that did crop up during the project fabrication, this made the design be limited in functionality.
2. The payload system was very expensive thereby making the installation of the camera, video transmitter and receiver expensive, hence it limited the project to using an android phone as an alternative mean incase the picture or video were to be taken.
3. During the flight test, at one time the quadcopter crashed hard from a height of around 20 meters while testing the battery capacity, the quadcopter charge dropped 7.5 Volts which is the minimum voltage that can maintain the quadcopter in air and hence resulting to that fatal fall that resulted to breaking all the propellers and one arm of the quadcopter .
4. Flying a quadcopter was the hardest challenge as needed a prior knowledge of flying a similar model of quadcopter of an airplane, these been my first time to make the quadcopter limited me to making crucial decision like taking off the quadcopter from the air and safe landing the quadcopter.

**Conclusion**

Though this was an ambitious project, a lot of preliminary designs were considered during the research in order to develop a versatile quadcopter that would serve as a tool to undertake Aerial Security Surveillance System. The anticipated results from the design of the quadcopter were as follows:
1. The designed quadcopter should weigh at least 1.5Kgs
2. The quadcopter should take off and land safely.
3. The quadcopter should take all commands given and interpret them effectively
4. The quadcopter should hover laterally and vertically with ease.

The prototype designed is user friendly and can be easily used to satisfy the specific goals outlined in chapter one, among all the above anticipated results the designed Quadcopter was able to take off and landing safely was achieved.

The quadcopter had a problem with constant lateral motion thus making control an issue that after doing the test analysis, the control and flying the quadcopter was a big challenge, lastly the quadcopter crashed several times before learning proper control and flying.

**Recommendations**

Despite success of this project, there is still work to be done, given more time and resources many changes would have been done in the design, additional sensors could also be added to improve the flight performance such as sonar and an optical flow sensor. These steps would help increase precision of flight (especially in wind), stability of the quadcopter, and the overall tracking. Also a camera could be installed in the quadcopter that can send video and picture to a base station that would receive information and take the necessary decision. In design to make the quadcopter full autonomy obstacle sensor is necessary feature, which could include a sonar sensor in front of the Quadcopter, this would be a simple and quick addition to prevent the quadcopter from running into objects, another alternative would be to use a camera or several cameras and implement a form of computer vision to detect objects and determine the best course to circumnavigate the object. Adding obstacle avoidance would drastically increase the usefulness and market-readiness of this prototype. Lastly the individual learning to fly quadcopter should seek proper guidance from the civil aviation department which as an institution charged with the training and certification of licensing of pilots.

**3.14 Cost Analysis.**

Table 3.4 Cost Analysis

| Material required | Specification and type | Quantity | price |
|---|---|---|---|
| Transmitter | 6-channel<br>4-channel | 1 | |
| Receiver | Model flysky CT6B radio<br>Specifications<br>Channel: 6 channel<br>Frequency 2.4 GHz<br>RF power less than 20db<br>Modulation: GFSK<br>Code type: PCM<br>Sensitivity: 1024<br>DSC port: yes<br>Power 12V (1.5AAA*8)<br>ANT length:26mm<br>Certificate CE FCC | 1 | **Ksh. 10,000 For both transmitter and receiver.** |
| Brushless motor | **Use** - Vehicles & Remote Control Toys<br>**For Vehicle Type -** Helicopter<br>**Material** – Metal<br>**RC Parts & Accs** – Motors<br>**Model Number –** A2212 13T 1000KV | **4** | **3616.72** |
| Electronic Speed Controller | Red Bricks 30a Esc<br>5v Power Supply | 4 | **1958.88** |
| Propellers And Prop Adapter Rings | Specifications<br>10x4.5-2 Blade Prop<br>10x6.3-3 Blade Prop | 8 | **517.28** |

| Lithium Polymer Batteries | Specifications<br>• 11.1 V 3 cell pack<br>• 2200mAh of charge<br>• 25C continuous discharge<br>• Size 103x34x15mm<br>• Weight 130g | 1 | **2538.70** |
|---|---|---|---|
| Flight Control Board | Model kk2.1 Board<br>Specifications<br>Size 50mmx50mmx23.5mm<br>Weight 14.5g<br>Ic At mega 328pa<br>Gyro: Murata Piezo<br>Input Voltage: 3.3-5v<br>Signals From Receiver:1520us (4channels)<br>Signal To Esc 1520us | 1 | **2734.80** |
| Connecting wires | A bunch | | **300** |
| Strip board | | 3 set | **200** |
| Bread board | | 3 set | **600** |
| Soldering gun | | 1 | **300** |
| Soldering wire | | 4mtrs | **200** |
| screws | | 1 box | **250** |
| **TOTAL** | | | **Ksh 23216.38** |

## 3.15 Project time management schedule

| Task/time | SEPT 2015 | OCT 2015 | NOV 2015 | DEC 2015 | JAN 2016 | FEB 2016 | MAR 2016 | APR 2016 | MAY 2016 |
|---|---|---|---|---|---|---|---|---|---|
| **Research** | ■ | ■ | ■ | ■ | ■ | ■ | | | |
| **Proposal and mini-presentation** | | | | | ■ | ■ | | | |
| **Project design** | | | ■ | ■ | ■ | ■ | ■ | ■ | |
| **Hardware component collection** | | | | ■ | ■ | ■ | ■ | ■ | |
| **Assembling circuit and testing** | | | | | ■ | ■ | ■ | ■ | |
| **Final report writing** | | | | | | ■ | ■ | ■ | ■ |
| **Final presentation** | | | | | | ■ | ■ | ■ | ■ |

# APPENDIX I

## MATLAB PROGRAM OF QUADCOPTER RESULTANT BLADE VELOCITY.

```
%clear
Vmph=28
rpm=1000
omega=rpm*2*3.14/60;
Vkt=Vmph/1.15077;
Radius=4/12;
V=Vkt/0.5925;
psi1=0:0.01:2*pi;
rx=Radius*cos(psi1);
ry=Radius*sin(psi1);
m=-Radius*.45;
y=0;
psi=0;
n=-25:1:25;
for i=1:1:100
y=y+Radius/100;
for j=1:1:100
psi=psi+2*pi/100;
U(i,j)=omega*y+V*sin(psi);
crx(i,j)=y*cos(psi);
cry(i,j)=y*sin(psi);
end
end
v=[-100 -50 0 50 100 200 300 400 500 600 700 800 900 1000 1100];
[C,h]=contour(crx,cry,U,v);
colormap cool
clabel(C,h)
hold
plot(rx,ry,'r')
plot(n,m)
axis square
xlabel('rotor radius (ft)')
ylabel('rotor radius (ft)')
title(['Quadcopter Resultant Blade Velocity in Forward Flight',' Velocity =',num2str(Vkt),'kts',' \Omega =' ,num2str(omega), 'rad/sec'])
```

# APPENDIX II
# MATLAB SIMULATION OF QUADCOPTER RESULTANT BLADE VELOCITY.

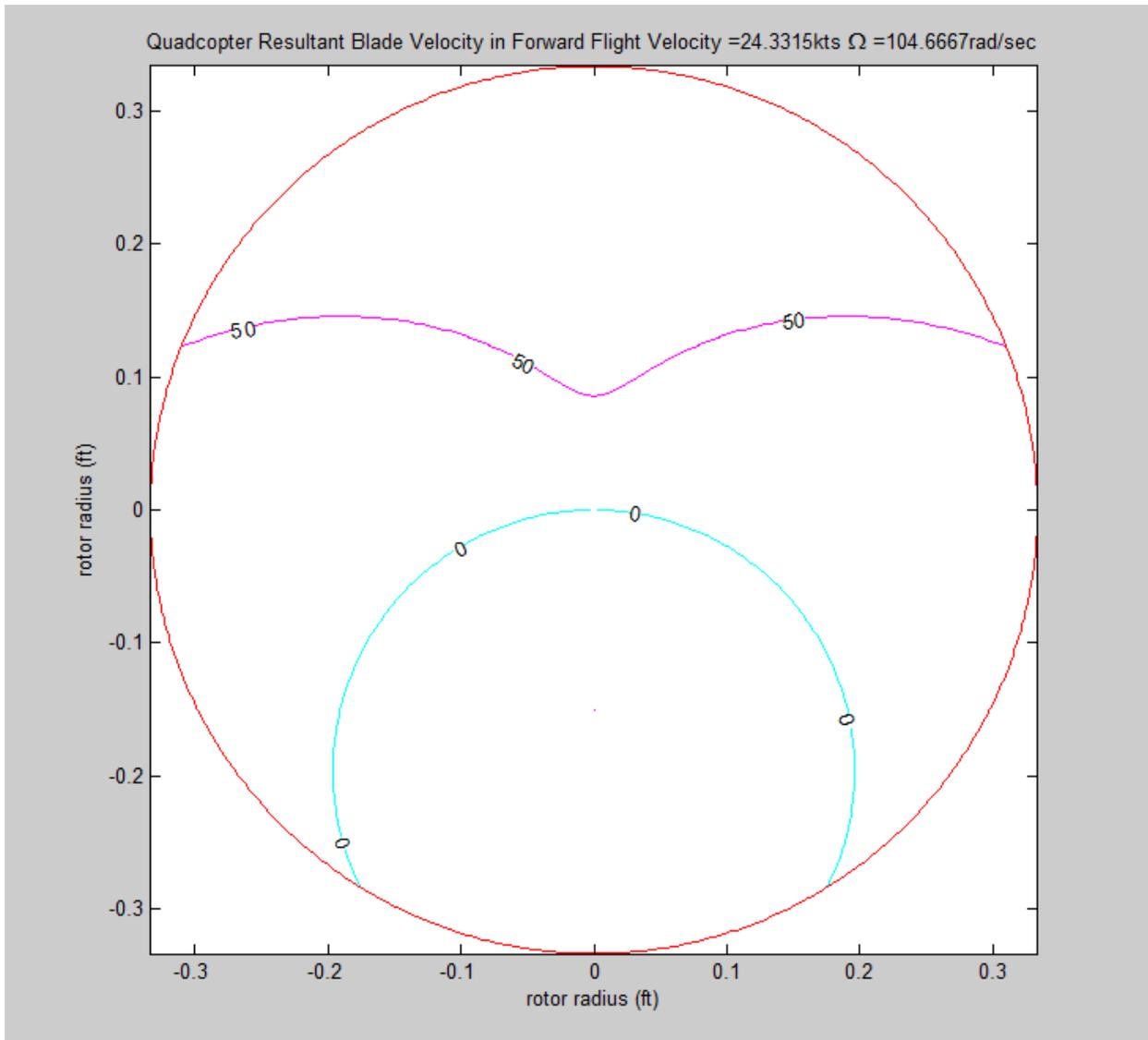

# APPENDIX III

```
tdata = csvread('asthl20log.csv',1,0);
ts = timeseries([convang(tdata(:,[3 2]),'deg','rad') ...
          tdata(:,4) convang(tdata(:,5:7),'deg','rad')],tdata(:,1));
h = Aero.FlightGearAnimation;
h.TimeseriesSourceType = 'Timeseries';
h.TimeseriesSource = ts;
h.FlightGearBaseDirectory = 'C:\Program Files\FlightGear191';
h.FlightGearVersion = '1.9.1';
h.GeometryModelName = 'HL20';
h.DestinationIpAddress = '127.0.0.1';
h.DestinationPort = '5502';
h.AirportId = 'KSFO';
h.RunwayId = '10L';
h.InitialAltitude = 7224;
h.InitialHeading = 113;
h.OffsetDistance = 4.72;
h.OffsetAzimuth = 0;
h.TimeScaling = 5;
get(h)
image(imread([matlabroot filesep fullfile('toolbox','aero','astdemos','figures','astfganim01.png')],'png'));
axis off;
set(gca,'Position',[ 0 0 1 1 ]);
```

# APPENDIX IV

**FLIGHT TRAJECTORY**

# APPENDIX VI

## CIRCUIT DIAGRAM

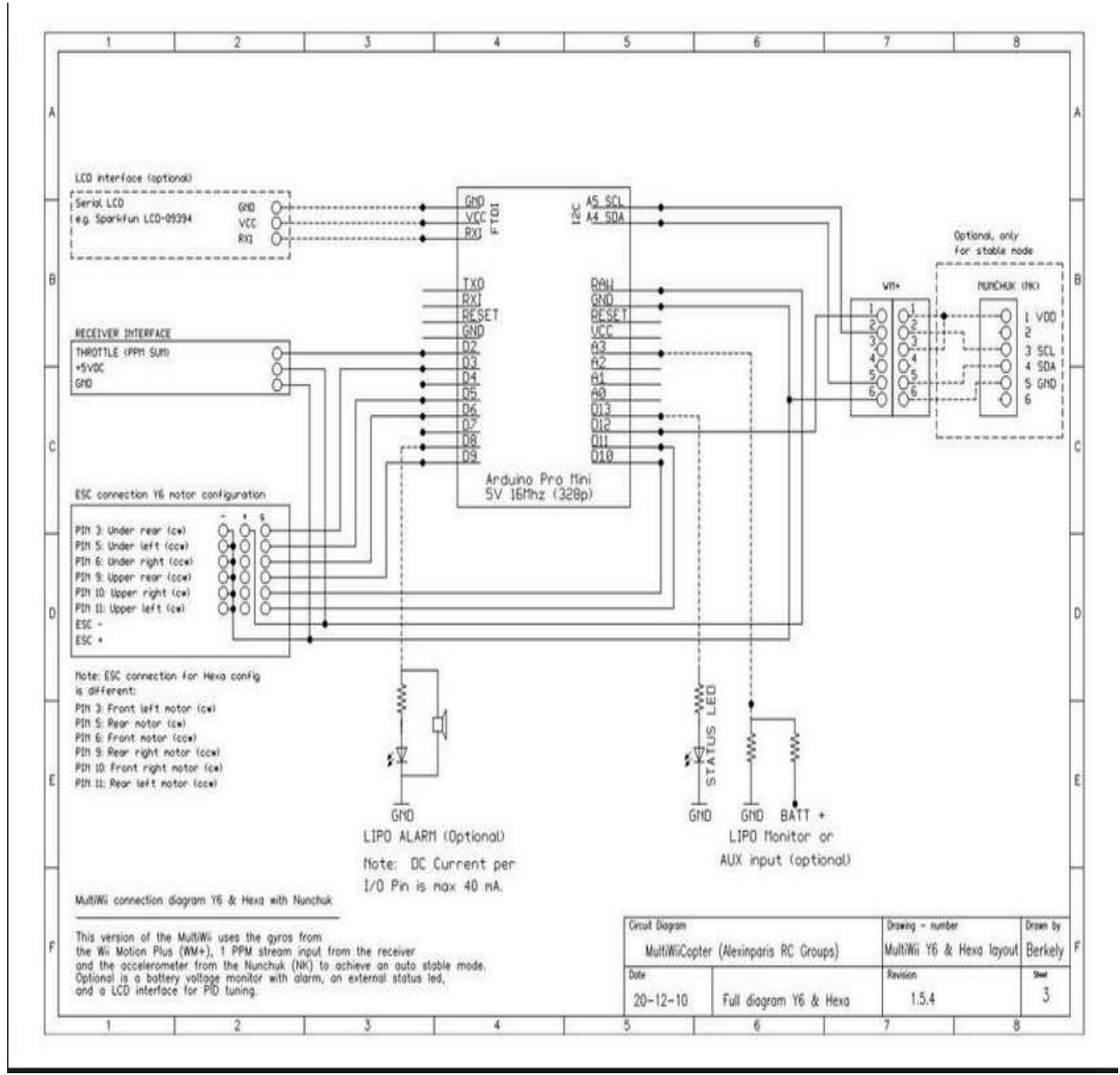